\documentclass[iop]{emulateapj}
\usepackage{hyperref}       % needed to use pdflatex and include graphics as PDF files only
\usepackage{color}
\newcommand{\Msun}{M_{\odot}}
\newcommand{\Mvir}{M_{\rm vir}}
\newcommand{\Rvir}{R_{\rm vir}}

\newcommand{\HI}{H$\,$I}
\newcommand{\enzo}{\texttt{Enzo}}
\newcommand{\ramses}{\texttt{Ramses}}
\newcommand{\art}{\texttt{Art}}
\newcommand{\arepo}{\texttt{Arepo}}
\newcommand{\gizmo}{\texttt{Gizmo-PSPH}}
\newcommand{\gasoline}{\texttt{Gasoline}}
\newcommand{\yt}{\texttt{yt}}

\submitted{}

\begin{document}
\title{High Angular Momentum Halo Gas: a Feedback and Code-Independent Prediction of LCDM}
\author{Kyle R. Stewart\altaffilmark{1},
        Ariyeh H. Maller\altaffilmark{2,3},
        Jose O{\~n}orbe\altaffilmark{4},
        James S. Bullock\altaffilmark{5,6},
        M. Ryan Joung\altaffilmark{7},
        Julien Devriendt\altaffilmark{8},
        Daniel Ceverino\altaffilmark{9},
        Du{\v s}an Kere{\v s}\altaffilmark{10},	
        Philip F. Hopkins\altaffilmark{11},
        and Claude-Andr{\' e} Faucher-Gigu{\` e}re\altaffilmark{12}
       }

\altaffiltext{1}{Department of Mathematical Sciences, California Baptist University, 8432 Magnolia Ave., Riverside, CA 92504, USA}
\altaffiltext{2}{Department of Physics, New York City College of Technology, 300 Jay St., Brooklyn, NY 11201, USA}
\altaffiltext{3}{Department of Astrophysics, American Museum of Natural History, Central Park West at 79th Street, New York, NY 10024, USA}
\altaffiltext{4}{Max-Planck-Institut für Astronomie, Königstuhl 17, D-69117 Heidelberg, Germany}
\altaffiltext{5}{Center for Cosmology, Department of Physics and Astronomy, The University of California at Irvine, Irvine, CA, 92697, USA}
\altaffiltext{6}{Center for Galaxy Evolution, Department of Physics and Astronomy, The University of California at Irvine, Irvine, CA, 92697, USA}
\altaffiltext{7}{Department of Astronomy, Columbia University, New York, NY 10027, USA}
\altaffiltext{8}{Department of Physics, University of Oxford, The Denys Wilkinson Building, Keble Road, Oxford OX1 3RH, UK}
\altaffiltext{9}{Zentrum für Astronomie der Universität Heidelberg, Institut für Theoretische Astrophysik, Albert-Ueberle-Str. 2, 69120 Heidelberg, Germany}
\altaffiltext{10}{Department of Physics, Center for Astrophysics and Space Sciences, University of California at San Diego, 9500 Gilman Drive,La Jolla, CA 92093, USA}
\altaffiltext{11}{California Institute of Technology, 1200 E. California Boulevard, Pasadena, CA 91125, USA}
\altaffiltext{12}{Department of Physics and Astronomy and CIERA, Northwestern University, 2145 Sheridan Road, Evanston, IL 60208, USA}

\begin{abstract} {
We investigate angular momentum acquisition in Milky Way-sized galaxies by comparing 
five high resolution zoom-in simulations,
each implementing identical cosmological initial conditions but utilizing different hydrodynamic codes: \enzo, \art, \ramses, 
\arepo, and \gizmo.  Each code implements a distinct set of feedback and star formation prescriptions.  
We find that while many galaxy and halo properties vary between the different codes (and feedback prescriptions), there is qualitative agreement on 
the process of angular momentum acquisition in the galaxy's halo.  In all simulations, cold filamentary gas accretion to the halo results in 
$\sim4$ times more specific angular momentum in cold halo gas ($\lambda_{\rm cold}\gtrsim0.1$) than in the dark matter halo.
At $z>1$, this inflow takes the form of inspiraling cold streams that are co-directional
in the halo of the galaxy and are fueled, aligned, and kinematically connected to filamentary gas infall along the cosmic web. 
Due to the qualitative agreement among disparate simulations, we conclude that the buildup of high angular momentum halo gas and 
the presence of these inspiraling cold streams
are robust predictions of Lambda Cold Dark Matter galaxy formation,
though the detailed morphology of these streams is significantly less certain.  
A growing body of observational evidence suggests that this process is borne out in the real universe. 
}
\end{abstract}
\keywords{galaxies:formation---galaxies:halos---galaxies:evolution --- methods:numerical}

\section{Introduction}
\label{Introduction}

In the standard Lambda Cold Dark Matter (LCDM) picture of galaxy formation, gas accreting onto a growing dark matter 
halo shock-heats to the virial temperature of the halo, giving the gas time to virialize and eventually cool out of 
the hot gaseous halo and sink into the central galaxy \citep{ReesOstriker77,Silk77,WhiteRees78,WhiteFrenk91,MallerBullock04}.
Under this picture of galaxy growth, it is expected that the resulting angular momentum distribution of galaxies should
mimic the spin of their dark matter, resulting in rotationally supported galaxy disks (and presumably hot gaseous halos as well) that 
are proportional to the spin of the dark matter halo \citep{FallEfst80,MoMaoWhite98}, which has been well studied in dissipationless
$N$-body simulations and semi-analytic merger trees \cite[e.g.][]{Bullock01, Vitvitska02, Maller02,Avila-Reese05,DOnghia07,Bett10,Munoz-Cuartas11,Ishiyama13,Trowland13,Kim(Choi)15}.

However, recent advances in hydrodynamic simulations and galaxy formation theory have increasingly emphasized the 
importance of ``cold flows''---gas accretion onto galaxy halos via filamentary streams with cooling times shorter than the 
compression time for establishing a stable shock\footnote{Some recent moving-mesh simulations have called into question whether these cold streams deliver unshocked 
gas \emph{to the galaxy} without heating in the inner regions of the halo \citep[e.g.][]{Torrey12,Nelson13,Nelson15,Nelson16}.  
As our focus in this work is on gas accretion into the \emph{halo}, not the 
eventual transition from the halo to the galaxy, this distinction should have minimal impact on the topics discussed here.}, 
either when the halo is below a critical mass threshold, or even for massive halos at sufficiently high redshift 
\cite[e.g.][]{Binney77,Keres05, DekelBirnboim06, Ocvirk08, Brooks09, Dekel09, FGKeres10, FG11, Stewart11a, vandeVoort11,Hobbs15,vandeVoort15}.
In the cold flow paradigm, gas that is accreted in the cold mode tends to have specific angular momentum considerably higher than the dark matter 
\citep{Chen03, SharmaSteinmetz05,Keres09,KeresHernquist09,Agertz09,Brook11, Stewart11b, Kimm11}, inconsistent with the previous picture of galaxy angular momentum 
buildup. The resulting angular momentum of the stellar disk may be rather different from that of the accreted gas because of feedback effects \citep{MallerDekel02,Brook11}.

As a result of this changing paradigm for cosmological gas accretion and galaxy growth \citep[for a recent review, see][]{Stewart17review}, 
a new scenario of angular momentum acquisition in galaxies and galaxy halos seems to be emerging.  
In this picture \citep{Stewart11b,Kimm11,Pichon11,Codis12,Danovich12,Stewart13, Codis15,Danovich15,Prieto15,Tillson15}
the particularly high angular momentum of cold flow gas is related to its coherent,
filamentary origin, coupled with the specific geometry of the cosmic web in the environment of a given galaxy.  
These filamentary cold flows deliver significant angular momentum to galaxy halos, 
with the cold gas orbiting for $\sim1-2$ dynamical times before spiraling into the central galaxy.  
At any given time, galaxy disks typically have lower spin than halo gas, owing to the fact that the specific angular momentum 
of infalling material increases with time.  Halo gas is ``younger'' and this correlates with higher spin.

Importantly, this scenario is predictive.  
The high-spin halo gas is often (but not always) coherent in its spin direction, 
with inspiraling cold streams often forming a thick planar structure of high angular momentum cool gas that co-rotates with the central disk.   
It is important to emphasize that while this extended gas tends to rotate, it is not angular-momentum supported.  
Rather, this gas usually spirals in on $\sim2$ dynamical times.  
Though not perfectly aligned with the orientation of the galactic disk, the inspiraling halo gas usually has coherent rotation along a preferred plane.

Encouragingly, an increasing number of observations have begun to demonstrate the abundance of high angular momentum material 
in galaxy halos, qualitatively consistent with this emerging theoretical picture.  In the local universe, some of these observations include
detection of high angular momentum extended \HI$ $ disks and XUV disks \citep{Oosterloo07, 
ChristleinZaritsky08,Sancisi08,Lemonias11,Holwerda12}, as well as low metallicity high angular momentum gas (presumably from fresh accretion) in polar ring galaxies \citep{Spavone10}.
There is even indication that local extended \HI$ $ disks may be environmentally dependent on the galaxy's filamentary environment \citep{Courtois15}.
At moderate redshift ($z\sim0.5$-$1.5$) numerous absorption line studies of the circumgalactic medium of galaxies have begun to emphasize the bimodal properties of absorbers, where 
absorption along the galaxy's major axis tends to show high angular momentum (co-rotating) inflow, and absorption along the galaxy's minor axis shows observational 
signatures of outflow \citep{Kacprzak10,Kacprzak12a,Kacprzak12b, Bouche12,Bouche13,Crighton13,Nielsen15,Diamond-Stanic15,Bouche16,Ho17}.
At higher redshift ($z\sim2$-$3$) kinematic studies of Ly$\alpha$ ``blobs'' have observed large-scale rotation consistent with high angular momentum cold gas accretion \citep{Martin14,Prescott15}.
There are also recent detections of massive protogalactic gaseous disks kinematically linked to gas inflow along a cosmic filaments, strikingly similar to the theoretical ``cold flow disk'' structure \citep{Martin15, Martin16}.

In this context, it is important that we ascertain how robust the predictions of these cosmological simulations are---a difficult task, considering that
many properties of simulated galaxies 
depend sensitively on the implementation of uncertain subgrid physics models such as gas cooling, star formation, radiation pressure, and supernova feedback
\cite[e.g.,][]{Thacker00,Kay02,Aquila,Gnedin11,PiontekSteinmetz11,Martizzi12,Agertz13,Vogelsberger13,Vogelsberger14,Marasco15,Genel15,Ceverino14,AgertzKravtsov16}.  
In addition, even with identical subgrid implementations, there are inherent numerical advantages and disadvantages between 
different hydrodynamic code implementations---for example, Lagrangian smoothed particle hydrodynamic (SPH) versus Eulerian grid 
codes---that result in artificial differences between galaxies simulated with different codes \cite[e.g.,][]{Morris96, Agertz07,Wadsley08,CullenDehnen10,Hahn10,Arepo,Hopkins15,Richardson16}. 

In order to test the validity of the emerging cold flow picture of angular momentum acquisition, we must ascertain the dependency of these predictions 
on the use of different numerical techniques and a variety of cutting-edge subgrid physics models.  
In this paper, we run five 
hydrodynamic zoom-in simulations of a Milky Way-sized galaxy, each with identical cosmological initial conditions but with different 
codes: \enzo, \ramses, \art, \arepo, and \gizmo, each implemented with recent subgrid physics models. 
In order to ensure uniform analysis for different hydrodynamic codes, we utilize the analysis software \yt$ $ \citep[which allows a single analysis routine 
to be run on different code architectures;][]{yt1} to explore the angular momentum content of halo gas and
whether or not the expected ``cold flow disk'' prediction is robust across these disparate platforms.  
We introduce the simulations in \S\ref{simulations}, present our main results from the comparison \S\ref{histories}-\S\ref{halo_scale}, finding that the same qualitative 
picture of high angular momentum halo gas in the form of co-directional inspiraling cold streams (which do occasionally  take the form of cold flow disks)
is present in all simulations---a seemingly natural consequence of filamentary gas accretion in LCDM.  
We discuss the implications of these results and the growing observational evidence of their existence
in \S\ref{discussion} and summarize and conclude in \S\ref{conclusion}.

%%---------------------------- Simulation Codes TABLE ---------------------
\renewcommand{\arraystretch}{1.5}
\begin{table*}[tbh!]
\begin{center}
\caption{SIMULATION CODE DETAILS}
\begin{tabular}{ | p{0.14\linewidth} | p{0.15\linewidth} | p{0.15\linewidth} | p{0.15\linewidth} | p{0.15\linewidth} | p{0.15\linewidth} | }
 \hline
                                     & \enzo & \art & \ramses & \arepo & \gizmo \\ \hline 
  Gravity Solver               &              FFT in the root grid & Multilevel particle mesh  & Multigrid particle mesh  & Tree multipole expansion particle mesh & Tree multipole expansion particle mesh  \\ \hline
  Hydrodynamics Solver &  Third-order piecewise parabolic method & Second-order Godunov method     & Second-order MUSCL scheme    & 2nd-order MUSCL scheme$^\dagger$ & Pressure-energy SPH \\ \hline
  High Res. $m_{\rm DM}^\ddagger$ &  $1.75\times10^{5}\Msun$  & $1.75\times10^{5}\Msun$  & $1.75\times10^{5}\Msun$  & $1.75\times10^{5}\Msun$  & $1.75\times10^{5}\Msun$  \\ \hline
  Grav.$\, \,$Softening [$h^{-1}$ comoving pc]   &   95 (DM, gas) & 95 (DM, gas)  & 95 (DM, gas)  & 95 (DM, gas) & 95 (DM), 14 (gas) \\ \hline
  SF Threshold            &   0.04 cm$^{-3}$ & 1 cm$^{-3}$ & 1 cm$^{-3}$ & 0.13 cm$^{-3}$ & 5 cm$^{-3}$ + self-grav. + molecular \\ \hline
  SF Efficiency             &   $\epsilon=0.03$ & $\epsilon=0.03$ & $\epsilon=0.03$ & $t_{SFR}=2.2$ Gyr & $\epsilon=1$ (in self-grav., molecular gas) \\ \hline
  Stellar Feedback$^\star$       &      Thermal & Thermal \& Rad. & Kinetic & Kinetic & Mixed [see text] \\ \hline
  Temperature Floor    &         10 K & 300 K & 100 K & 500 K & 10 K  \\ \hline
  UV Background         &         HM96 (increased Gaussian width) & HM96  & HM96 & FG09 & FG09 \\ \hline
  Reionization              &    $z=6$ & $z=7$ & $z=10$ & $z=10$  & $z=10$ \\ \hline
 
  \end{tabular}
  \vspace{-1 em}
  \tablenotetext{0}{\tablenotemark{} HM96 --- \cite{HaardtMadau96} } 
 \tablenotetext{0}{\tablenotemark{} FG09 --- \cite{FaucherGiguere2009} } 
 \tablenotetext{0}{\tablenotemark{$\dagger$} Subsequent versions of $\arepo$ have switched to a different time integration \citep{Pakmor16} using Heun’s method}
 \tablenotetext{0}{\tablenotemark{$\ddagger$} For Lagrangian codes, high resolution gas particle mass is $3.3\times10^{4}\Msun$}
 \tablenotetext{0}{\tablenotemark{$\star$} See text for detailed descriptions of feedback models.}
 %Enzo density threshold for SF is really 7e-26 g/cm^3, but converted to 1/cm^3, assuming pure hydorgen gas, for sake of comparison  -> 7e-26 / m_proton
  \end{center}
\end{table*}
\label{table_sims}
\vspace{1.3 em}
%%---------------------------- Simulation Codes TABLE ---------------------

\section{The Simulations}
\label{simulations}

\subsection{Overview}
The simulations used in this paper are all part of the Scylla Multi-Code Comparison Project. %\footnote{https://sites.google.com/site/scyllasims/} 
This project resimulates a Milky Way halo mass zoom-in simulation \citep[originally performed by Ryan Joung with \enzo$ $ in][]{Joung12} using other cosmological hydrodynamic codes.
While we focus on the redshift range $1<z<3$ in this work, we note that the resulting disk-type galaxy and its gaseous halo have already been 
studied in detail at low redshift \citep[e.g.,][]{Joung12,Fernandez12}.  Of particular importance to this work, \cite{Fernandez12} determined 
that the mass (in \HI$ $), covering fraction, and spatial distribution of the cold gas halo at $z=0$ are consistent with existing 
observations of nearby spiral galaxies. 

The codes are all run with their recent\footnote{Inevitably, however, there are bound to be further improvements to 
some of the subgrid models during the time it took to run the simulations, analyze, and publish the results.} 
subgrid models in order to compare state of the art simulations across codes.  Thus, the project is much like the Aquilla code
 comparison \citep{Aquila} but with higher resolution.  Our resolution is similar to the Agora code comparison project \citep{Kim14}, but that project is 
 seeking to use uniform physics while we are running each code as it has been used for other science papers.  
The codes used here are \enzo, \art, \ramses, \arepo$ $ and \gizmo. For all runs, the cosmology, dark matter particle mass, and box size are identical---the
box is 25 Mpc/h across with a much smaller region simulated at high resolution, using dark matter particles of mass $1.75 \times 10^5 \Msun$.  
All adaptive mesh refinement (AMR) codes reach the same maximum refinement of 95 $h^{-1}$ comoving pc, which is identical to the force resolution of the Lagrangian codes,
with the exception of gas particles in \gizmo, which uses an adaptive gravitational softening with a minimum value of 14 $h^{-1}$ comoving pc.
A flat cosmology consistent with WMAP5 \citep{WMAP5} is used throughout, with 
$\Omega_m = 0.279$, $\Omega_{\Lambda} = 0.721$, $\Omega_b = 0.046$, $h=0.70$, $\sigma_8 = 0.82$, and $n_s = 0.96$. 

The initial conditions for the original \enzo$ $ run \citep{Joung12} were generated with the code \texttt{Grafic}\footnote{http://web.mit.edu/edbert/} \citep{Grafic} with a starting redshift of $z=99$. 
The same code with the same seed was used to generate the initial conditions for the \ramses$ $ run. 
For all other runs, the dark matter particles from the \enzo$ $ run were used to determine the initial conditions. 
That is, the dark matter particles were set identical to those in the \enzo$ $ run and baryons were added based on the dark matter distribution 
(no separate transfer function).  We expect these differences to be negligible by the redshift where galaxies are forming.

Based on the cosmological model specified above, all Lagrangian codes set the gas mass resolution ($m_{\rm gas}=3.3 \times 10^4 \Msun$) 
relative to the dark matter particle mass ($m_{\rm DM} = 1.75 \times 10^{5} \Msun$).
Table \ref{table_sims} outlines many of the pertinent details for each code, including star formation (SF) density thresholds and efficiency parameters, 
epoch of reionization, UV background model, and the type
of stellar feedback model adopted.  Below, we describe the gas cooling and feedback physics of each individual run in more detail and include references to recently published science papers that 
utilize similar subgrid physics models as those implemented here.

For all analyses that follow, we make the distinction between ``cold'' and ``hot'' gas by a temperature cutoff of $250,000$ K 
\cite[commonly used as the distinction between ``cold-mode'' and ``hot-mode'' gas accretion; e.g.,][]{Keres05,Keres09,Stewart11b,Stewart13}.  
Using the Rockstar halo finder \citep{Rockstar} we calculate the virial radius of each simulation at each output redshift, 
finding that the halo-finding algorithms produce slightly different 
virial radii  at the same redshift for different simulations ($\pm5\%$ from the mean).
For the sake of identical comparison between codes, we therefore utilize a fitting function for $\Rvir(z)$ that averages over all simulations to adopt an
identical estimated virial radius for all simulations (at a given redshift), which varies from $68$ physical kpc at $z=3$ to $171$ physical kpc at $z=1$.

In order to guarantee uniform analysis for the varied code architectures and file formats, all analyses presented here have been performed utilizing the 
\yt\footnote{http://yt-project.org} analysis software \citep{yt1,yt2,yt3}, an open source project that has been developed and is continually being 
maintained and improved by the astrophysical community for the intended purpose of supporting cross-code compatible hydrodynamic analysis routines. 
In plots showing images of the gas distribution, we use a slightly older version of yt, (version yt-3.2.3) because after that version yt updated the way gas 
particles are deposited into cells.  Prior to the update, yt used clouds-in-cells deposition to determine the gas properties of a cell, while after the 
update the sph smoothing kernel is used instead. Although this update gives more accurate deposition for sph particles, it is very inaccurate for Arepo 
where the particle size relates to the volume of the cells from the Voronoi tessellation and not a smoothing length. The clouds-in-cells deposition gives 
adequate results for both methods so we use it for images.  For quantitative analysis, we use the gas particles for both \arepo$ $ and \gizmo$ $ so that no 
deposition into cells is required. 

%Enzo
\subsection{Enzo}
The \enzo$ $ \citep{Enzo} run serves as the basis for the Scylla simulation suite, and was performed in 2010 by Ryan Joung and discussed in \cite{Joung12, Fernandez12}; and \cite{Putman12}.  
\enzo$ $ uses an AMR grid to solve the equations of hydrodynamics, with this particular run 
using a version of \enzo$ $ before the uniform release of \enzo$ $ 2.0. 
$\enzo$ uses an FFT in the root grid gravity solver and a 3rd order piecewise parabolic method hydrodynamics solver.
Feedback is thermal as described in \citet{Cen05}.  
The simulation includes metallicity-dependent cooling to a temperature of $10$ K \citep{DalgarnoMcCray72}, neutral hydrogen shielding from UV radiation, 
and diffuse photoelectric heating \citep{Abbott82,  Joung09}.

%Art
\subsection{Art}
The \art$ $ \citep{Art1,Art2} run 
uses an AMR grid to solve the equations of hydrodynamics.  
$\art$ uses a multilevel particle mesh gravity solver and a 2nd order Godunov method hydrodynamics solver.  
Our run uses the star formation and feedback models described in \cite{Ceverino14} and 
includes thermal feedback from supernova explosions and stellar winds \citep{Ceverino09, Ceverino10}
as well as radiative feedback \citep[model $RadPre\_LS\_IR$ in][]{Ceverino14}.  
This model of radiative feedback includes radiation pressure from ionizing and infrared photons, photoheating, and 
photoionization from massive stars.
Other recent papers using similar physics include \cite{Zolotov15, Snyder15a, Ceverino15, Goerdt15, Mandelker17, Tacchella16a, Tacchella16b, Ceverino16, Tomassetti16} and \cite{Ceverino17}.

%Ramses
\subsection{Ramses}
The \ramses$ $ \citep{Ramses} run (\ramses$ $ version 3.0) uses
an AMR grid to solve the equations of hydrodynamics.
$\ramses$ uses a particle mesh gravity solver and a 2nd order MUSCL scheme hydrodynamics solver.
The gas cooling is based on a metallicity-dependent cooling, including metal
line cooling down to a temperature floor of $100$ K.  A stiffening of the interstellar medium (ISM) equation of state (chosen as a power law with $\gamma=4/3$) was
used to prevent gas with densities higher than the 1 atom/cm$^{-3}$ threshold to cool further than $100$ K and artificially fragment.
Feedback includes energy from stellar winds and supernovae (deposited in kinetic form) following \cite{DuboisTeyssier08}, where the proper distributions of SN II lifetimes are 
based on \cite{Starburst99} and \cite{Leitherer10}, such that energy from SNe II is injected continuously between $2$ and $50$ Myr.  Feedback from
SNe Ia are also included, following \cite{GreggioRenzini83} to compute the SN frequency.  This run has essentially the same physics as 
in \cite{Dubois14,Welker14,Codis15}; and \cite{Chisari15} with the exception that we have not included any AGN physics here.

%Arepo
\subsection{Arepo}
The Arepo \citep{Arepo} run 
employs a quasi-Lagrangian finite volume method for solving the hydrodynamic equations of motion \citep{Vogelsberger13}. 
The version of $\arepo$ used here employs a tree multipole expansion gravity solver and 
a 2nd order Godunov method hydrodynamics solver with a MUSCL scheme; however subsequent versions of Arepo have switched to a different time integration in the 
hydrodynamics solver \citep{Pakmor16} using Heun’s method.
Radiative gas cooling includes both primordial cooling \citep{Katz96} as well as line cooling from heavy 
elements \citep{Wiersma2009a, Vogelsberger13}.
Pressurization of the ISM, star formation, and associated feedback is handled using the \cite{SpringelHernquist03} subgrid model. 
Time delayed stellar mass return and metal enrichment is carried out \citep{Wiersma2009b, Vogelsberger13},
and kinetic star formation driven winds are employed 
with a wind velocity scaled to the local dark matter velocity dispersion.  
Winds are launched carrying 40\% of the local ISM metallicity to prevent over ejecting metal mass from the dense ISM \citep{Zahid2014}.

This run includes a physics implementation that is similar to that used in the \texttt{Illustris} simulation \citep{Vogelsberger14,Genel14} with the notable difference 
that no AGN physics is included here.  Other recent work that contains similar physics include 
\cite{Torrey2014, Wellons2015a, Wellons2015b, Torrey2015a, Torrey2015b, Snyder2015b, RodriguezGomez2015, Sales2015, Bray2015}; and \cite{Mistani2015}.

%>>>>>>>>>>>>>>>>>>>>>>>>>>>>>>>>>2Rvir density and temperature plots at z=3<<<<<<<<<<<<<<<<<<<<<<<<<<<<<<<<<
\begin{figure*}[tb!]
 \includegraphics[width=1.0\textwidth]{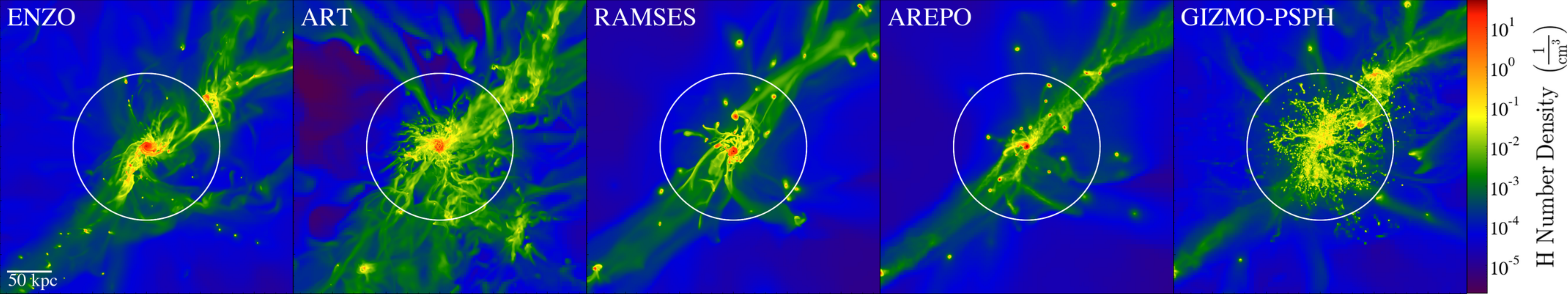}
 \includegraphics[width=1.0\textwidth]{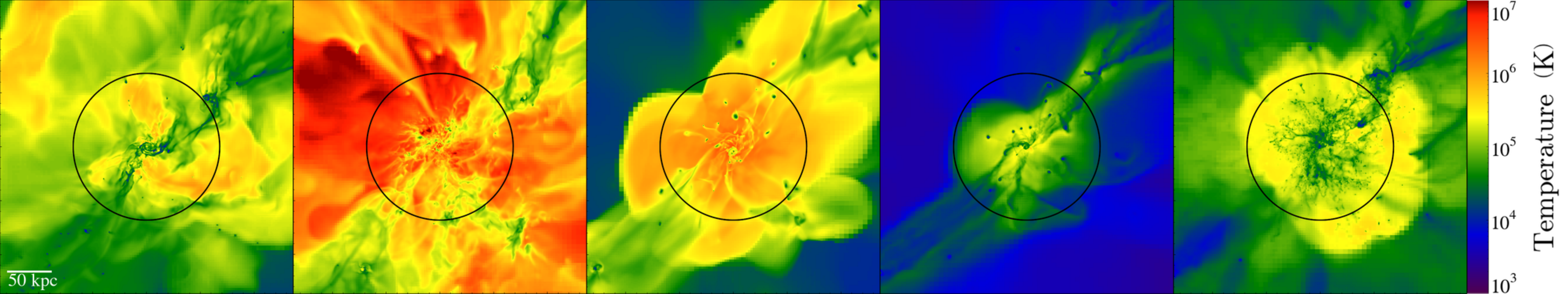}
 \caption{Hydrogen number density and temperature comparison at $z=3$.  All panels
   show the circumgalactic region, with panel widths of 272 physical kpc ($4\Rvir$).  Physical size scales are given in the leftmost panels, 
  and circles denote the virial radius of the halo.
  Top: projected gas density, showing the gas accretion onto the galaxy via cosmic filaments.  Detailed morphology of the resulting galaxy varies
  among simulation codes, but the same filamentary accretion structure is apparent.  
  Bottom: projected density-weighted gas temperature.  The temperature of gas in the CGM is highly dependent on the specific feedback and 
  code implementation.    
   }
\label{fig_z3_density_temp}
\end{figure*}
%>>>>>>>>>>>>>>>>>>>>>>>>>>>>>>>>>>>>>>>><<<<<<<<<<<<<<<<<<<<<<<<<<<<<<<<<

%Gizmo
\subsection{Gizmo-PSPH}
The $\gizmo$ \citep{Hopkins15} run uses a tree multipole expansion for the gravity solver and the pressure-energy formulation of smoothed 
particle hydrodynamics \citep[PSPH;][]{Hopkins13} together with a number of additional improvements to artificial viscosity, timestepping, and 
higher-order kernels, to solve the equations of hydrodynamics.\footnote{\texttt{Gizmo} is a multi-methods code that gives the user the choice of several hydrodynamic methods. 
This is why we use the label $\gizmo$ throughout this work, to distinguish the PSPH implementation from alternate methods.}
Radiative gas cooling includes both primordial cooling~\citep{Katz96} as well as cooling from 11 separately tracked metal species \citep{Wiersma2009a}. 
Gas follows an ionized + atomic + molecular cooling curve from $T=10-10^{10}$\,K. 

Star formation and feedback uses the Feedback In Realistic Environments (FIRE) prescriptions from \citet{Hopkins14}, which explicitly follow the mass, metal, momentum, 
and energy deposition by radiation pressure, photoionization and photoelectric heating, stellar winds, and SNe (Types II and Ia), with all rates tabulated from the 
stellar population model {\small STARBURST99} \citep{Starburst99} assuming a \citet{Kroupa01} IMF. They do not include AGN feedback. Unlike the other codes here, which assume stars form with a 
relatively low efficiency per free-fall time in all gas above some relatively large density threshold $\sim 0.1-1$ cm$^{-3}$, the FIRE models restrict star 
formation {\em only} to gas that is locally self-gravitating \citep[following][]{Hopkins13b}, self-shielding and molecular \citep[following][]{KrumholzGnedin11}, 
Jeans-unstable, and exceeds a higher density $n>5$ cm$^{-3}$, but within this highly restricted gas assumes the star formation occurs on a free-fall time. 
Other recent work that contains identical FIRE code and methods include \citet{Onorbe15, Chan15, Ma15,FG15} and \citet{Wheeler15}.

%>>>>>>>>>>>>>>>>>>>>>>>>>>>>>>>>>mass and SFR and spin vs time <<<<<<<<<<<<<<<<<<<<<<<<<<<<<<<<<
\begin{figure*}[t!]
 \includegraphics[width=0.33\textwidth]{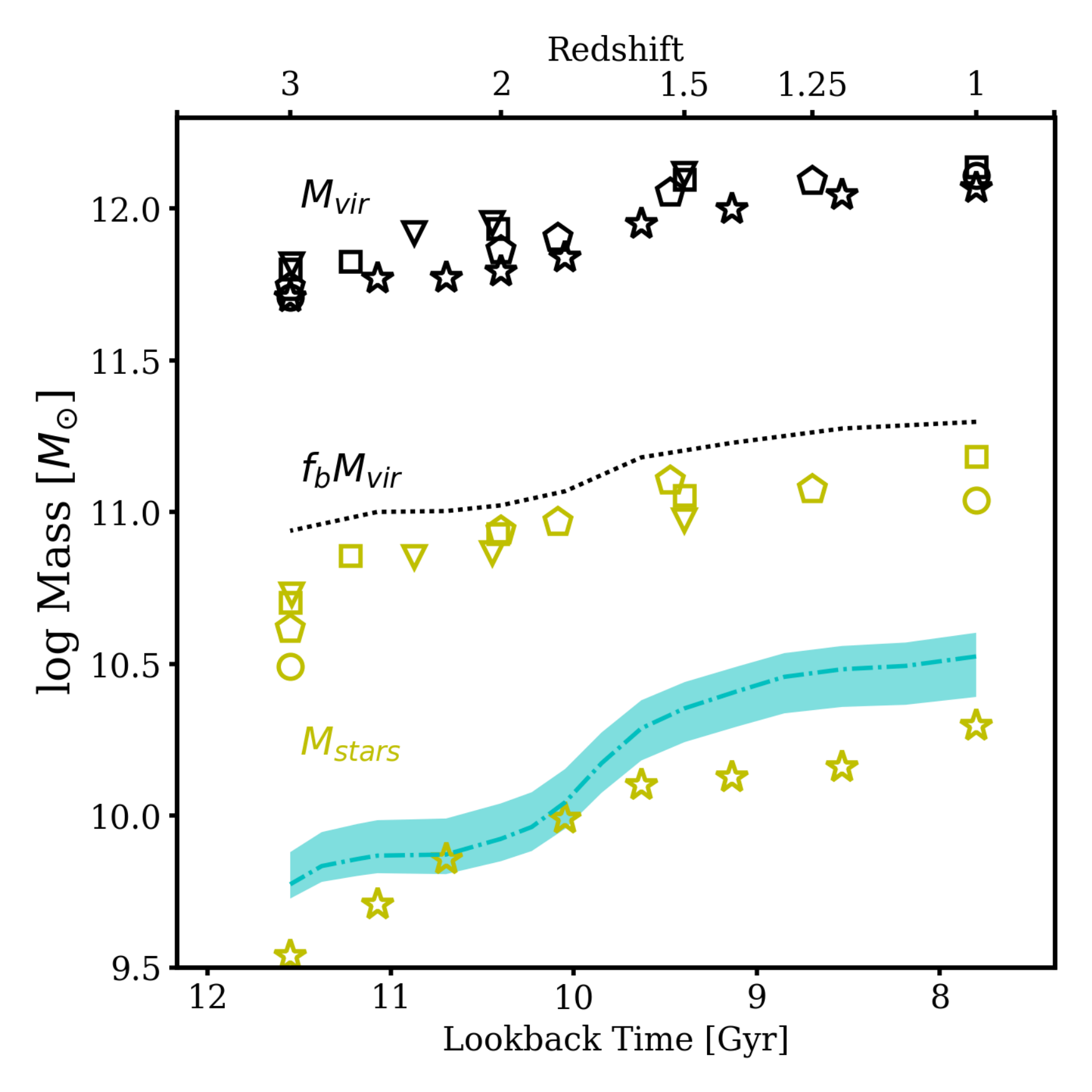}
 \includegraphics[width=0.33\textwidth]{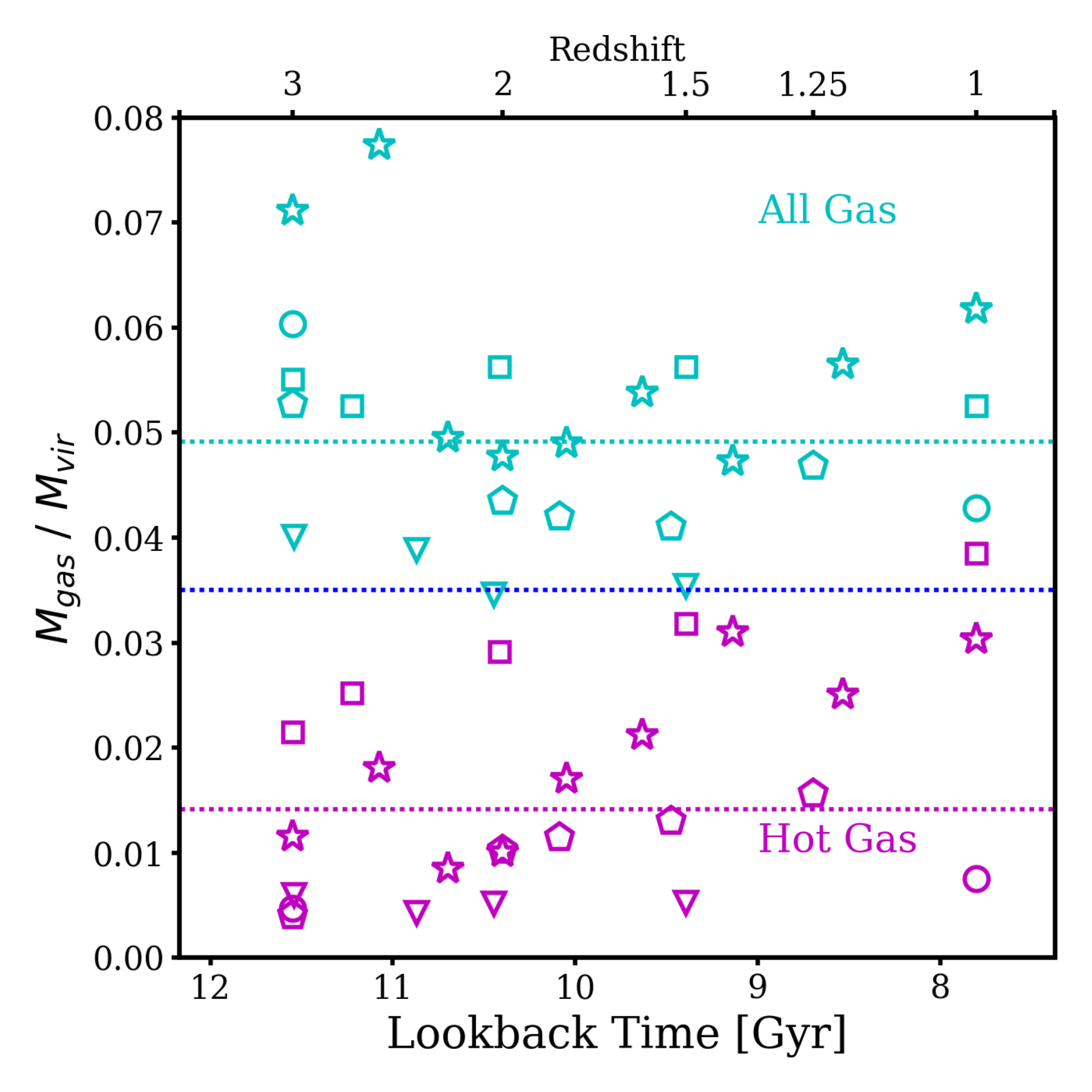}
 \includegraphics[width=0.33\textwidth]{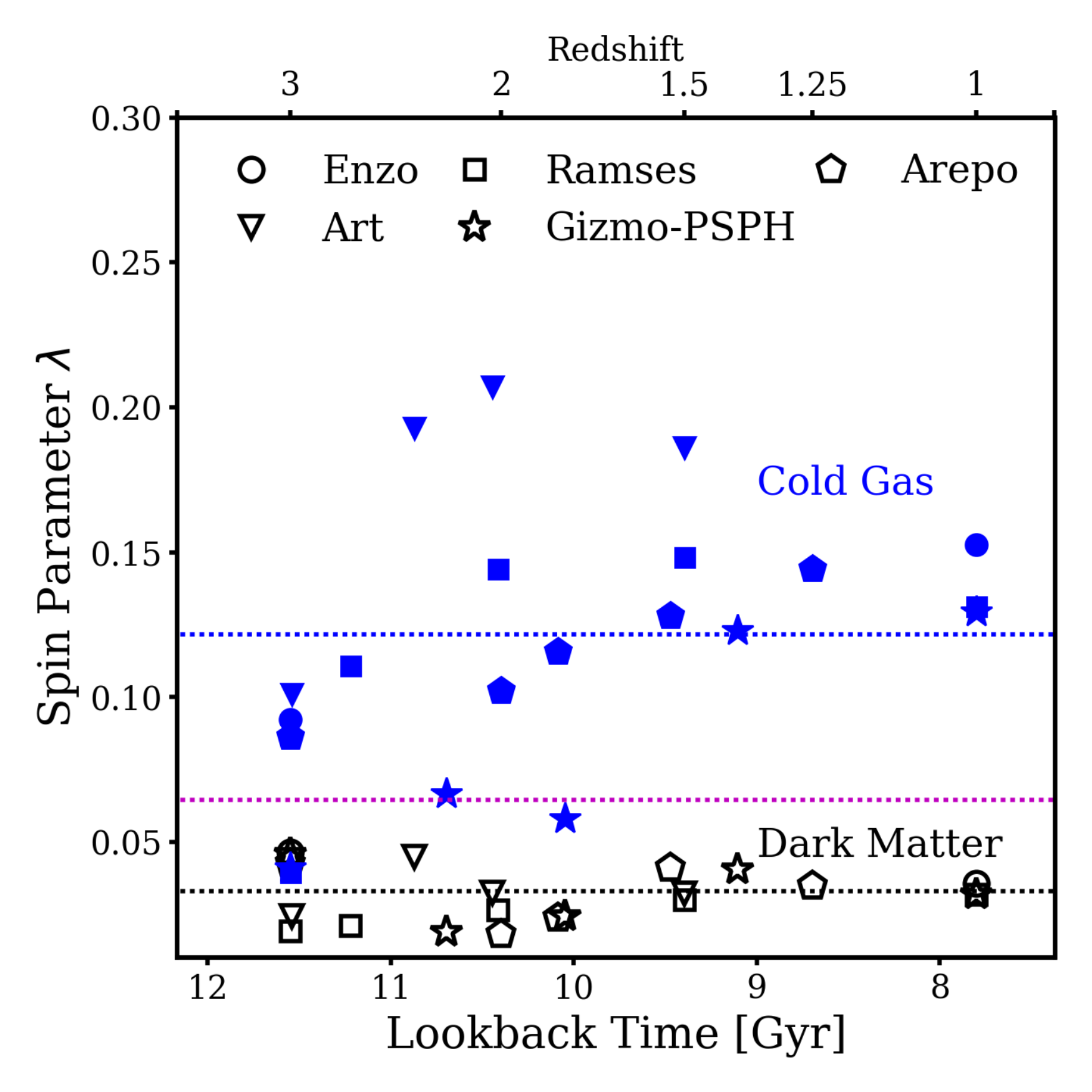}
\caption{
 Left: virial mass and galaxy stellar mass as a function of time from $z=3$-$1$.  Note that the total virial masses among simulations are quite similar; however, the stellar mass varies significantly.
                 Expectations from abundance matching are shown for comparison \citep[cyan shaded region:][]{Behroozi13}. 
  Middle: total gas fraction and \emph{hot} gas fraction in the halo (cold gas fraction not shown for the sake of clarity), vary noticeably depending on which code and feedback implementations 
           are used.  
  Right: spin parameter of cold halo gas and dark matter inside the galaxy halo (but excluding material within $R<0.1\Rvir$ so as not to include the galaxy).  
          Different symbols represent outputs from different simulations, and the mean values across all simulations for the cold halo gas, hot halo gas (symbols not shown for clarity) and 
            dark matter are given by the blue, magenta, and black horizontal dotted lines, respectively.
          All simulations demonstrate that cold halo gas has a significantly higher spin parameter compared to the dark matter, with typical values of $\lambda_{\rm cold}\simeq 0.12$.\
          For both the middle and right panels, the dotted horizontal lines represent averages over the entire redshift range and across all simulations for 
          all gas (cyan), cold gas (blue), hot gas (red), and/or dark matter (black).
   }
\label{fig_mass_vstime}
\label{fig_spin_histogram}
\end{figure*}
%>>>>>>>>>>>>>>>>>>>>>>>>>>>>>>>>>>>>>>>><<<<<<<<<<<<<<<<<<<<<<<<<<<<<<<<<

\section{Basic Halo Properties}
\label{histories}
\subsection{Large-scale Structure and Mass Growth}
We begin with a visual inspection of the region around the galaxy for each simulation.  Figure \ref{fig_z3_density_temp}
shows the gas density (number density of H; top) and density-weighted temperature (bottom) projections at $z=3$ through a cube of width $272$ physical kpc ($4\Rvir$ at this redshift).
The top panels of Figure \ref{fig_z3_density_temp} show qualitative agreement between the simulations on the general geometry and structure of the 
forming disk galaxy at this redshift, as well as its placement in a large-scale cosmic filament that is continually delivering an inflow of 
cold gas into the virial radius of the galaxy.  However, the detailed structure of the galaxy---and even that of the filament into 
which the galaxy is embedded---does appear to vary significantly between simulations.  For example, the width of the cosmic filament, the size and structure of the galactic disk, 
and the peak density of infalling satellite galaxies all vary on a noticeable level.

Perhaps more striking is the temperature differences among simulations shown in the bottom panels of Figure \ref{fig_z3_density_temp}.  All simulations demonstrate the presence of 
a significant gaseous halo around the galaxy, as well as streams of filamentary gas that penetrate the halo and deposit cold gas in the inner halo, near the 
galactic region.  However, the extent that feedback has 
enriched the CGM and IGM, the density structure of the gaseous halo, the temperature distribution of hot gas, and the precise structure of the cold flows as they interact with the 
gaseous halo of the galaxy vary significantly among simulations.

To illustrate some of the similarities and differences among the simulations,
the two left panels of Figure \ref{fig_mass_vstime} show the mass growth of the halo as a function of time, including the total mass (black) and galaxy stellar mass (yellow) 
on the leftmost panel, as well as the cold (blue) and hot (magenta) gas fractions  
within the virial radius.  
Note that the total virial mass (left panel) is quite similar among simulations, despite very different feedback implementations.
Comparing the linear scale of the middle panel to the log scale of the left panel, we also note that the gas fractions (middle panel)
are relatively similar among the simulations, although there are still noticeable variations.  
The average total / cold / hot gas fractions during the entire redshift range $z=3-1$ averaged over all the simulations is represented by the horizontal dotted lines in the figure.
Not surprisingly (given the mass scale of this halo), all simulations show that the dominant supply of halo gas is in a cold phase, 
rather than a massive reservoir of hot gas. 

The total galaxy stellar mass (left panel) shows a much more significant variation among simulations, with the 
$\gizmo$ code in particular forming a much smaller stellar mass than any of the other codes used here, likely as a result of 
strong feedback implementations.  
As a comparison, the upper limit of baryonic mass (the virial mass times the cosmic baryon fraction) 
is shown here as a thin black dotted line, and the shaded cyan region shows the expected galaxy stellar mass range for the given virial mass 
based on abundance matching \citep[median value $\pm1\sigma$ from][]{Behroozi13}.
Interestingly, the galaxy stellar mass for $\gizmo$ is much closer to observational expectations, and may even be slightly underproducing stars in the simulated galaxy, rather than overproducing them, as in the other simulations.   
As our goal in this work is to focus on \emph{similarities} between codes, with emphasis on the galaxy halos and not the galaxies themselves, 
we defer a more detailed discussion of the numerous differences between the simulations and their implications for galaxy formation as a topic for future study.

\subsection{Angular Momentum}
One fundamental result of the recent emerging picture of angular momentum acquisition in galaxies is that 
gas in the halos of galaxies tends to have specific angular momentum $\sim3-5$ times higher 
than the dark matter \citep{Stewart11b, Kimm11, Stewart13,Danovich15}.
We revisit these previous findings by comparing the spin parameter, $\lambda$, of both the 
cold halo gas and the dark matter in the halo for all our simulations.  
We adopt the spin parameter from \cite{Bullock01}: $\lambda_x \equiv j_x/\sqrt{2}VR$, where $\lambda_x$ is the spin parameter of a given component, based on that component's 
specific angular momentum, $j_x$, and $V$ and $R$ are defined by the virial velocity and virial radius of the halo, respectively.  
The right panel of Figure \ref{fig_spin_histogram} shows the spin parameter for each saved output of each simulation between $z=3$ to $z=1$, where we only include material inside the virial radius but 
outside of the central region ($0.1<R/\Rvir<1.0$) in our calculations since we are interested in the \emph{halo}, not the galaxy itself.
While the simulations vary in the precise value (and direction---not shown) of the angular momentum of their gaseous halos, we find 
several important qualitative agreements across all the simulations.

\begin{enumerate}
   \item Cold halo gas (and hot halo gas---not shown in the figure, for clarity) consistently has more specific angular momentum than the dark matter component. 
   \item While simulations agree that dark matter halo spin parameters are typically $\lambda_{\rm DM}\sim 0.03$, 
     the average cold halo gas spin parameter across our simulations is significantly higher: $\lambda_{\rm cold}\simeq0.12$.
   \item In agreement with previous work, averaging over all simulations, the cold halo gas contains $\simeq 4$ times the specific angular momentum of the dark matter halo (though with considerable variation), 
     while the hot gas typically has $\simeq 2$ times the specific angular momentum of the dark matter.
\end{enumerate}

%>>>>>>>>>>>>>>>>>>>>>>>>>>>>>>>>>Environments large scale z=3<<<<<<<<<<<<<<<<<<<<<<<<<<<<<<<<<
\begin{figure*}[tb!]
 \includegraphics[width=1.0\textwidth]{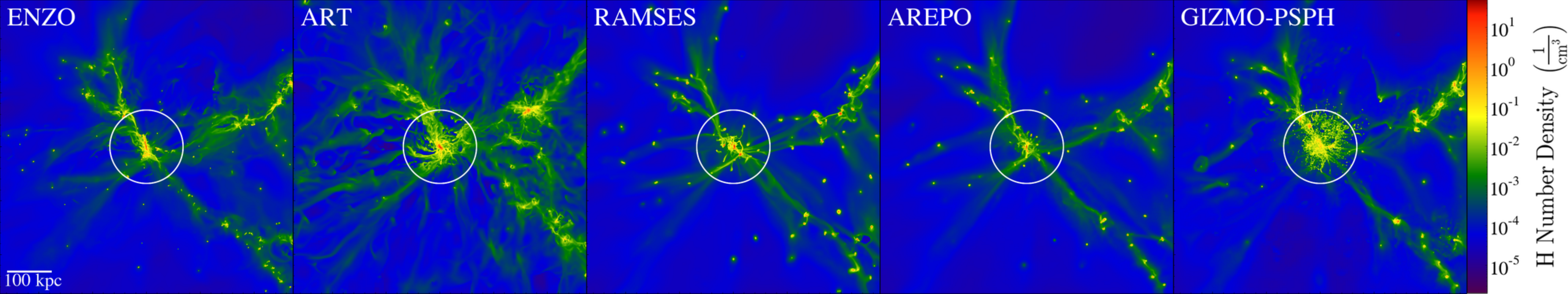}
 \includegraphics[width=1.0\textwidth]{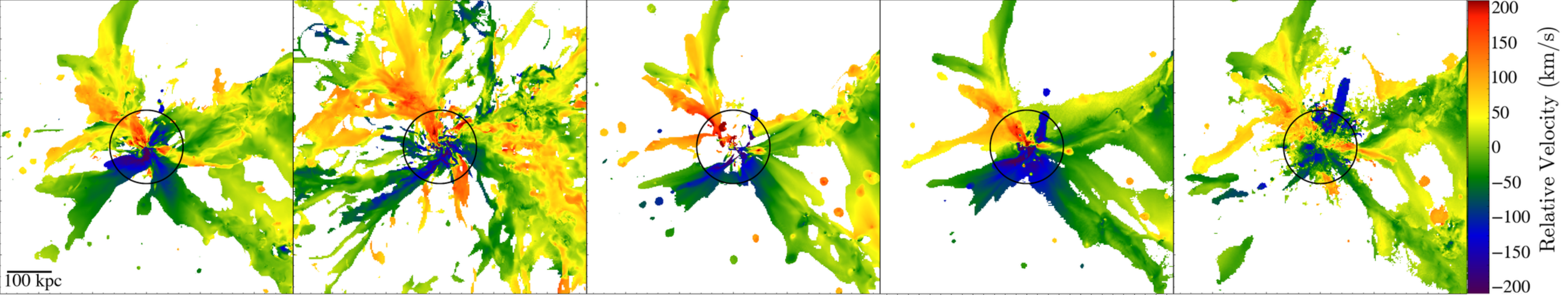}
 \caption{Large-scale environment at $z=3$ (along an orthogonal line of sight from Figure \ref{fig_z3_density_temp}).  
  In all panels, circles denote the virial radius of the halo and the size scale in physical kpc is indicated in the leftmost panels, 
  with panel widths of $8\Rvir$ ($544$ physical kpc). 
  Top: gas density projections showing the structure of the cosmic web near the galaxy.  
  Bot: line-of-sight velocity of cold gas with a density in hydrogen of $n_{\rm H}>3\times10^{-4}$ cm$^{-3}$ in an identical orientation and scale to the top panels. 
  There is a clear line-of-sight velocity signature---the top-left filament is redshifted while the bottom filaments are blueshifted---indicating 
  the motion of the cosmic web as it flows onto the massive galaxy halo.  
  (The particularly chaotic structure of the \gizmo$ $ simulation is due to a violent merger-induced outflow at this epoch; see \S\ref{coldflow_regrow}.)  
   }
\label{fig_z3environment}
\end{figure*}
%>>>>>>>>>>>>>>>>>>>>>>>>>>>>>>>>>>>>>>>><<<<<<<<<<<<<<<<<<<<<<<<<<<<<<<<<

These findings confirm previous results: the angular momentum of galaxy halos varies significantly among components;
the dark matter invariably measures a cumulative combination of past accretion, resulting in the lowest specific angular momentum; 
the hot gaseous halo is typically built and maintained 
both by non-filamentary ``hot-mode'' gas accretion, as well as feedback and outflows (which are sensitive to subgrid physics models); 
and the cold halo gas traces filamentary ``cold-mode'' accretion and has the highest specific angular momentum \citep{Stewart13}.
Thus, while our simulations agree with previous $N$-body simulations for a dark matter halo spin parameter, one should expect to observe 
typical cold halo gas with significantly higher angular momentum, with spin parameters $\lambda_{\rm cold}\sim0.1$.  

In a previous study of four cosmological zoom-in simulations (all using the same hydrodynamic code),
\cite{Stewart13} found no significant trend between cold gas spin parameter and cosmological time 
(at least, not significant enough to be 
apparent with a non-statistical sample of high resolution zoom-in simulations).  
Therefore, while Figure \ref{fig_spin_histogram} arguably shows a trend of increasing cold gas spin parameter from $z=3-1$, this may be a 
consequence of this particular halo's unique merger and accretion history, and is not likely to be a general result of galaxy formation in LCDM.

\section{Large-scale Filamentary Inflow}
\label{environment_scale}
In order to place this high angular momentum cold halo gas in the proper cosmological context, 
Figure \ref{fig_z3environment} shows the large-scale environment around the simulated galaxy---where, 
for purposes of this work, we define the halo environment by box widths of $8\Rvir$ (544 physical kpc at $z=3$).  
The top panels again show the gas density (H number density), similar to Figure \ref{fig_z3_density_temp} but zoomed out by a factor of two and viewed along an orthogonal orientation. 
The bottom panels show the line-of-sight velocity of all 
cold gas above a minimum density threshold in hydrogen (all forms) of 
$n_{\rm H}>3\times10^{-4}$ cm$^{-3}$, which was  
chosen to select only gas sufficiently dense to be embedded in filamentary (or dark matter halo) structures on these large-scale environments.

Because the galaxy is the most massive halo in its environment (i.e.~not a 
member of a group or cluster), the cosmic filaments in its environment are strongly affected by the halo potential, with
gas, dark matter, and smaller galaxies all flowing along the filaments toward the galaxy, 
demonstrated by the clear line-of-sight velocity indications in the bottom panels.  
For example, the filament to the upper left of the galaxy (situated in front of the galaxy along this line-of-sight) 
consistently shows redshifted velocities in all simulations, 
while the two filaments below (and behind) the galaxy are consistently blueshifted.  

This result is perhaps not surprising, as any three-dimensional filamentary structure where matter flows along the 
cosmic web toward a central overdensity (and is viewed along an arbitrary axis) is unlikely to show multiple filaments all flowing perpendicular to the line-of-sight. 
Thus, one should naively expect strong line-of-sight velocities to be apparent when viewing large-scale filamentary gas flows.
Although this may not be a \emph{surprising} result, it is important to keep these large-scale gas flows in mind for future discussion of the kinematics  
of inspiraling cold streams in the galaxy's halo.  We will see in \S\ref{halo_scale} that these large-scale filamentary flows have a direct impact on the behavior of the cold gas 
within the virial radius of the halo.  

Note that the line-of-sight velocity structure of the filament flowing in from the \emph{right} of the galaxy 
shows considerably more variation between the simulations.  This occurs because this filament \emph{does} happen to be roughly 
perpendicular to the line-of-sight.  Thus, the velocities along the rightmost filament are more sensitive to the 
peculiar velocities of galaxies, gas streams, and outflows, which vary more strongly between simulations
than the gross large-scale flows toward the central halo.

%>>>>>>>>>>>>>>>>>>>>>>>>>>>>>>>>>Halo Scale at z=3 <<<<<<<<<<<<<<<<<<<<<<<<<<<<<<<<<
\begin{figure*}[tb!]
 \includegraphics[width=1.00\textwidth]{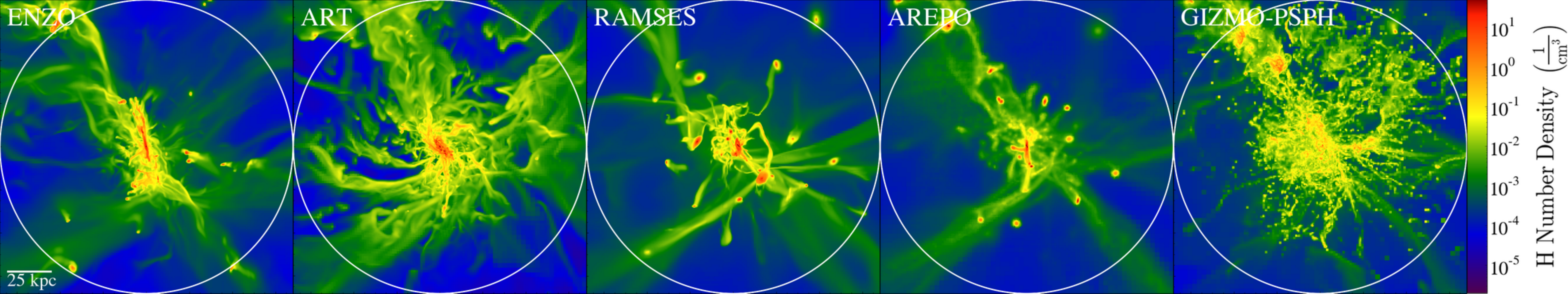}
 \includegraphics[width=1.00\textwidth]{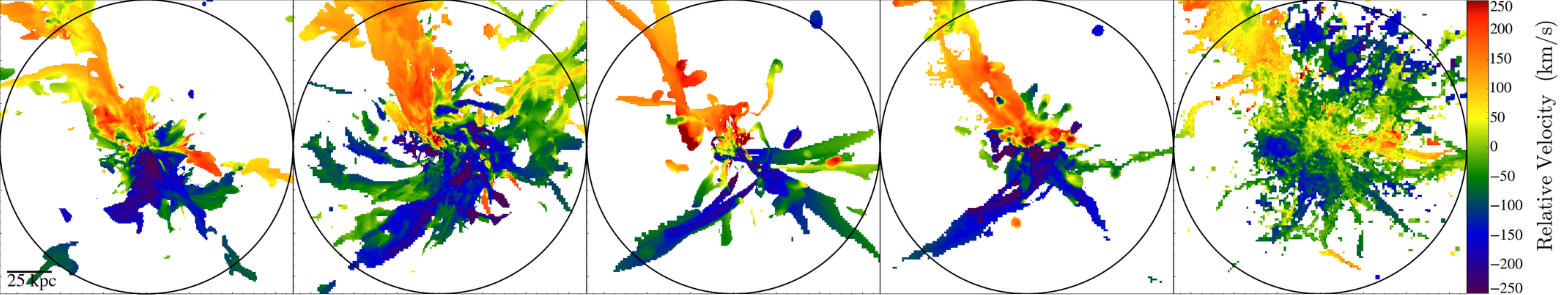}
 \includegraphics[width=0.98\textwidth]{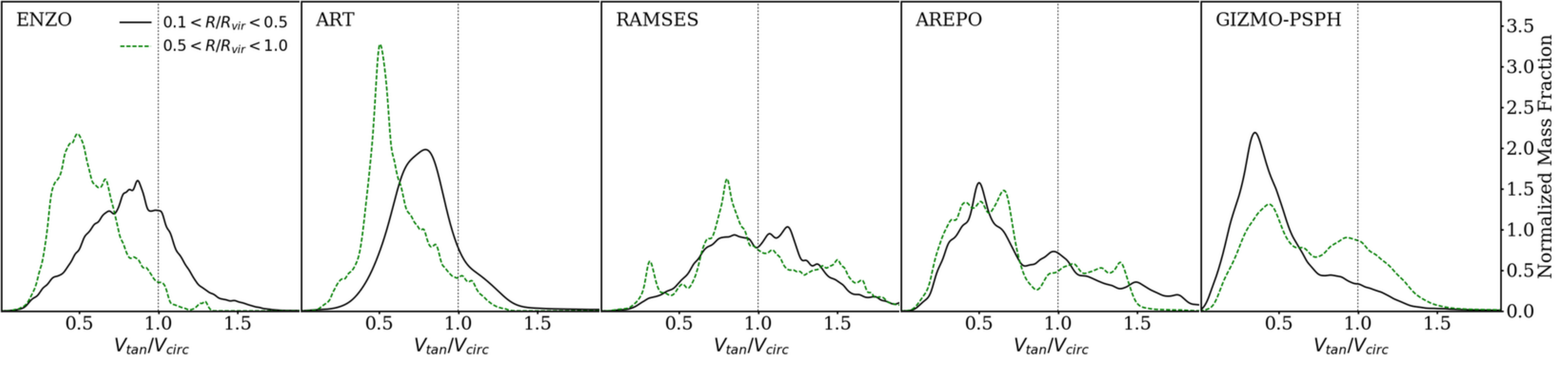}
 \caption{
  Top/middle: density projections and line-of-sight velocities at $z=3$, similar to Figure \ref{fig_z3environment} except that panels have now been 
  ``zoomed in'' to the virial radius of the halo, 
  and the minimum density threshold for hydrogen gas in the bottom panels has been increased by a factor of $10$ from Figure \ref{fig_z3environment} to 
  $n_{\rm H}>3\times10^{-3}$ cm$^{-3}$ (corresponding to column densities of $N_{\rm HI}\gtrsim10^{17}$ cm$^{-2}$). 
  Circles denote the virial radius of the halo and the size scale in physical kpc is indicated in the leftmost panels. 
  The kinematics of the co-directional inspiraling cold streams appear linked to the large-scale filaments that are 
  fueling them, as seen in Figure \ref{fig_z3environment}.
  (See \S\ref{coldflow_regrow} regarding the chaotic structure of the \gizmo$ $ simulation.)
  Bottom: $V_{\rm tan}/V_{\rm circ}$ at radius $R/\Rvir$ as a function of $R/\Rvir$ for cold dense gas in the halo.
  Most cold dense gas in the halo spirals in toward the center of the halo, and is not angular-momentum supported.
   }
\label{fig_coldflowdisks}
\end{figure*}
%>>>>>>>>>>>>>>>>>>>>>>>>>>>>>>>>>>>>>>>><<<<<<<<<<<<<<<<<<<<<<<<<<<<<<<<<

\section{Inspiraling Cold Streams}
\label{halo_scale}
We begin investigating the morphology and kinematics of cold halo gas in Figure \ref{fig_coldflowdisks}, which is analogous to Figure \ref{fig_z3environment},
except that the panels now focus only on material within the virial radius (panel widths of $136$ physical kpc at $z=3$). 
The bottom panels again show line-of-sight velocity maps 
of cold dense gas\footnote{We select gas based on temperature and density rather 
than \HI$ $ content or species column density because we want to avoid any differences in ionization fractions among simulations when making our comparison.
The qualitative trend that there is always high-angular momentum inspiraling gas in the halo
does not depend on the details of this selection criterion, though quantitative measures (e.g., the apparent covering 
fraction of this gas) will of course depend on these details---a topic we plan to revisit in future work.}
in the halo, except that we have increased the minimum density threshold by a factor of $10$ when compared to Figure \ref{fig_z3environment}, to
a hydrogen density of $n_{\rm H}>3\times10^{-3}$ cm$^{-3}$ \citep[this should correspond to a minimum hydrogen column density of $N_{\rm HI}\gtrsim10^{17}$ cm$^{-2}$,][]{Altay11, Schaye01}. 

The exact morphology of gas in the halo of the galaxy varies considerably among the simulations, which is not surprising, given the vastly 
different feedback mechanisms implemented in each simulation, some of which drive explosive spherical outflows that violently 
shred the ISM and CGM of the galaxy (e.g., $\gizmo$)
and some of which instead drive high-velocity bi-conical outflows out of the plane of the galaxy (e.g., \enzo).  
However, we also note that 
some of the morphological differences may also be influenced by the precise timing of galaxy mergers.  For example, the $\gizmo$ simulation
is in the midst of a violent outflow at this epoch, due to a recent merger, which partially explains the significantly more chaotic structure shown in Figure \ref{fig_coldflowdisks}
(we will demonstrate in \S\ref{coldflow_regrow} that the \gizmo$ $ simulation's line-of-sight velocity structure is much more similar to the other simulations 
immediately before \emph{and after} this merger-driven outflow event).
While the same general merger and accretion history takes place
for each simulation, the exact timing of these mergers at a given epoch may vary, and any coherent velocity structure for cold gas 
in the galaxy's halo is typically destroyed during a sufficiently strong outflow event.

%>>>>>>>>>>>>>>>>>>>>>>>>>>>>>>>>>Environments large scale z=4,2,1<<<<<<<<<<<<<<<<<<<<<<<<<<<<<<<<<
\begin{figure*}[th!]
 \includegraphics[width=0.48\textwidth]{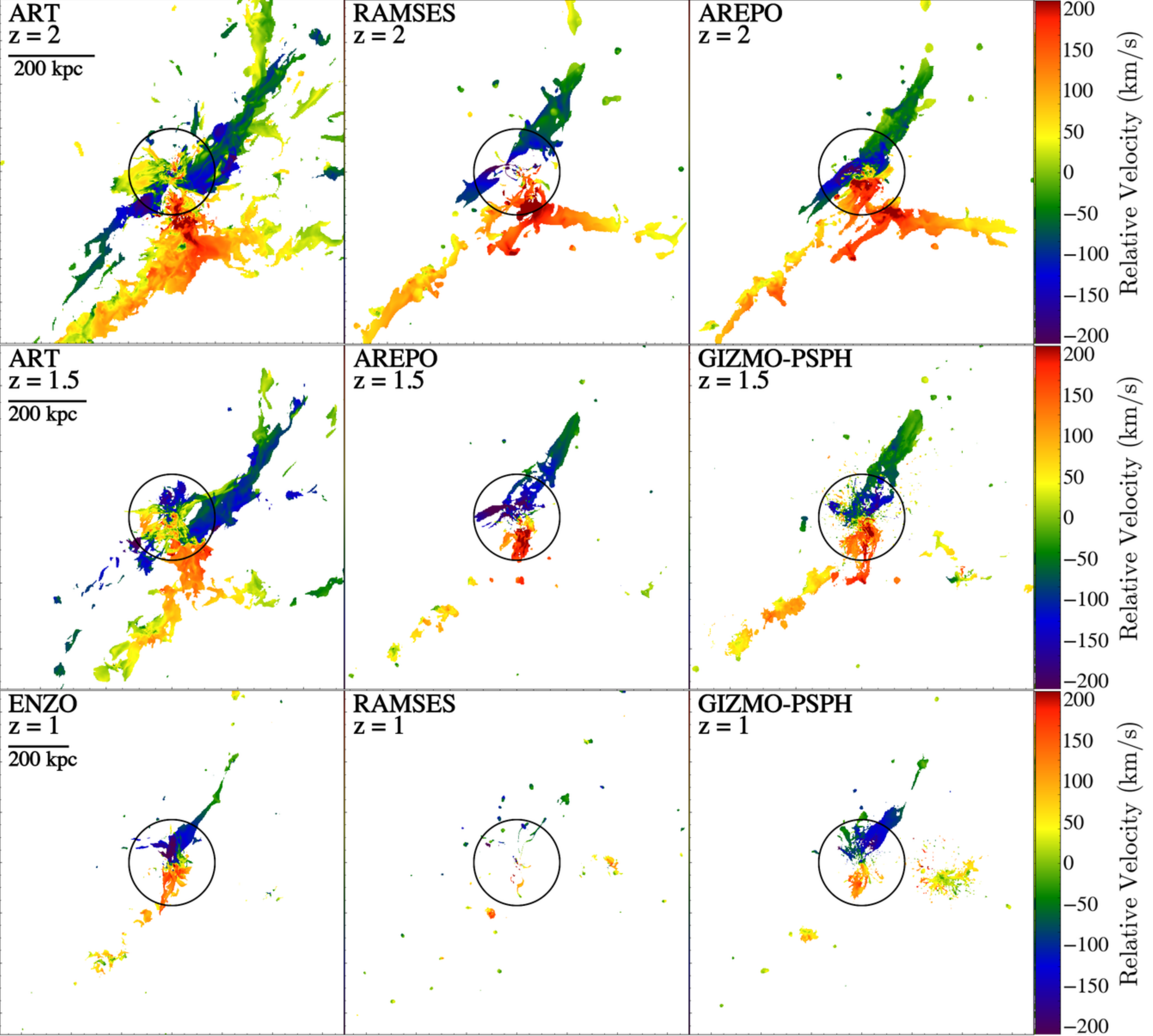}
\hspace{1em} 
\includegraphics[width=0.48\textwidth]{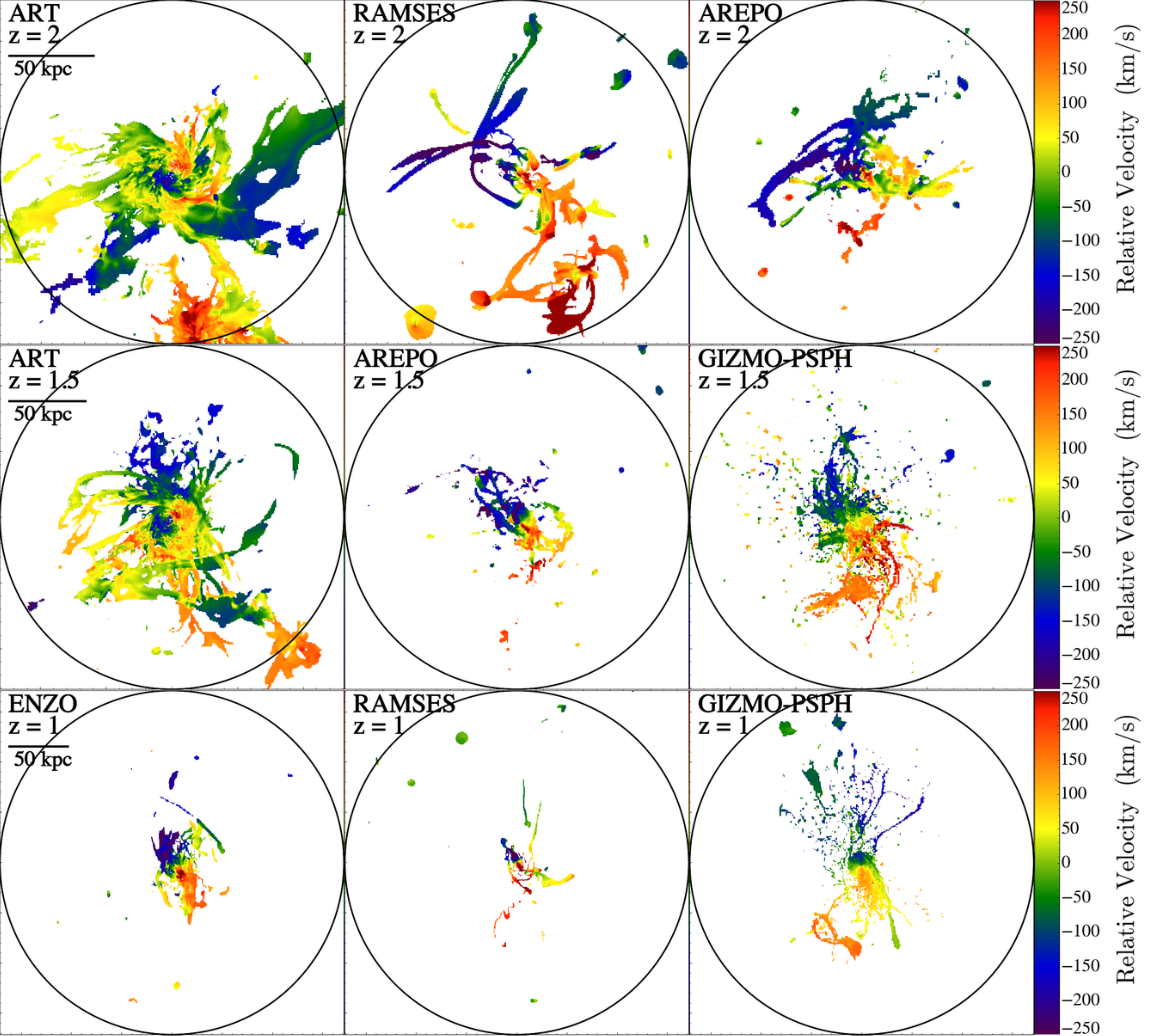}
 \caption{Left: large-scale environments at $z=2$ (top), $z=1.5$ (middle), and $z=1$ (bottom) for various subsets of the simulation runs.  
  As with Figure \ref{fig_z3environment}, the panels show the line-of-sight velocities of dense gas with hydrogen density of $n_{\rm H}>3\times10^{-4}$ cm$^{-3}.$
  Each box has a width of $8\Rvir$ ($\sim 0.8, 1.0, 1.4$ physical Mpc at $z=2, 1.5, 1$, respectively).
  Right: analogous  line-of-sight velocities, but zoomed-in to the virial radius (similar to Figure \ref{fig_coldflowdisks}), and with 
  an increased density threshold: $n_{\rm H}>3\times10^{-3}$ cm$^{-3}$ (corresponding to $N_{\rm HI}\gtrsim10^{17}$ cm$^{-2}$).  
  In all panels, circles denote the virial radius of the halo and the size scale in physical kpc is indicated in the leftmost panels of each row. 
  Note the clear signature of inspiraling cold streams, kinematically linked to the large-scale filamentary gas flowing onto the galaxy.}
\label{fig_z21environment}
\label{fig_coldflowdisks2}
\end{figure*}
%>>>>>>>>>>>>>>>>>>>>>>>>>>>>>>>>>>>>>>>><<<<<<<<<<<<<<<<<<<<<<<<<<<<<<<<<

Despite these varied differences in morphology, the middle panels of Figure \ref{fig_coldflowdisks} show a similar qualitative picture. 
As was the case with the large-scale environment,
the cold gas entering the virial radius from the upper-left filament shows a dramatic redshift in each simulation, while the cold gas entering from the bottom 
filaments show strong blueshifts.  
(As before, the line-of-sight velocity of the material in the upper-right quadrant of these panels is less uniform, as it probes a gas accreting along a filament that is roughly 
perpendicular to the line-of-sight.) 

The bottom panels of Figure \ref{fig_coldflowdisks} show that while the angular momentum content of the halo gas is high, the bulk of this inflowing gas 
does not have enough angular momentum to be fully rotationally supported. 
That is, most---but certainly not all---of the cold dense halo gas at $R>0.1\Rvir$ has a tangential velocity $V_{\rm tan}/V_{\rm circ}<1$ 
(where $V_{\rm circ}=\sqrt{GM(<R)/R}$ is the circular velocity at a given radius).
Thus, despite the clear velocity structure shown in the middle panels of the figure, this high angular momentum gas should not be considered rotationally supported, but rather 
spiraling in toward the center of the halo, consistent  with the short ``sinking times'' of $\sim1-2$ halo dynamical times previously reported by \cite{Stewart11b}.

The qualitative result in each case is 
a clear co-directional inflow signature, with cold dense halo gas easily divided by a single cutting plane into the 
redshifted versus blueshifted half of the halo, flowing through the halo via a chaotic assortment of high angular momentum 
inspiraling cold streams that are kinematically linked to inflow from the cosmic web.

Similar structures have been noted a number of times in the literature, but with a variety of terminologies, including the  ``messy region'' \citep{Ceverino10}, ``cold flow disks'' \citep{Stewart11b, Stewart13},
the ``AM sphere'' \citep{Danovich12}, or ``extended rings'' \citep{Danovich15}.  Indeed, depending on the simulation code utilized, one can easily see in Figure \ref{fig_coldflowdisks} 
how the kinematics and morphology of the 
inspiraling streams may or may not be well-described as a ``messy region'' (e.g. \ramses) or a more orderly disk-like structure (e.g. \art).  
Thus, while we find that the exact morphology---including size, orientation, clumpiness, thickness---of any structure that results from the inspiraling cold streams 
may vary significantly among simulations,  
each code does produces a qualitatively similar picture in which there is a clear line-of-sight velocity structure within the virial radius of the halo that is kinematically 
linked to that of the large-scale filamentary environment of Figure \ref{fig_z3environment} (with the exception of the \gizmo$ $ simulation at this epoch; see \S\ref{coldflow_regrow}).

Figure \ref{fig_z21environment} shows line-of-sight velocity maps (along an orthogonal orientation) for dense cold gas at $z=2$ (top), $z=1.5$ (middle), and $z=1$ 
(bottom) for various subsets of the simulation runs (as labeled).  
The left panels of this figure look at the large-scale environment (analogous to Figure \ref{fig_z3environment}.  
While the basic filamentary nature of the gaseous inflows becomes less apparent at 
decreasing redshift (when the filaments are less dense), we can still note the 
same qualitative behavior of inflowing gas.  On environmental scales, filamentary inflow results in the same clear
line-of-sight velocity signature as before; across all simulations, gas flowing into the virial radius from the top of the panels is blueshifted, 
while gas flowing in from the bottom is redshifted, with the only notable exception being the $\ramses$ code at $z=1$, 
which is likely the result of the lack of self-shielding from the UV background, 
leaving very little cold gas above our minimum density threshold, so almost no cold dense inflow is still visible in the figure.
While the detailed structure of the inflowing gas again varies among simulations, it seems apparent that 
filamentary gas accretion along a three-dimensional cosmic web 
onto an overdense region (at this mass scale in the redshift range $1<z<3$) tends to produce the same qualitative picture across
all the simulations, regardless of the subgrid physics.

The right panels of Figure  \ref{fig_coldflowdisks2} shows an analogous line-of-sight velocity analysis, but zoomed-in to the 
halo virial radius for $z=2$ (top), $z=1.5$ (middle), and $z=1$ (bottom), and again increasing the density threshold by a factor of $10$ (similar to Figure \ref{fig_coldflowdisks}).
Again, the precise structure of the inspiraling cold streams varies among the simulations, but 
most of the simulations produce qualitatively similar pictures; there continues to be a clear large-scale velocity structure within the virial radius of the halo that is kinematically 
linked to the large-scale filamentary inflow shown in the left panels, again with the exception of $\ramses$ at $z=1$, which has evacuated most of its halo of cold dense gas altogether.

%>>>>>>>>>>>>>>>>>>>>>>>>>>>>>>>>>Co-directional mass fraction <<<<<<<<<<<<<<<<<<<<<<<<<<<<<<<<<
\begin{figure*}[tb!]
 \includegraphics[width=0.32\textwidth]{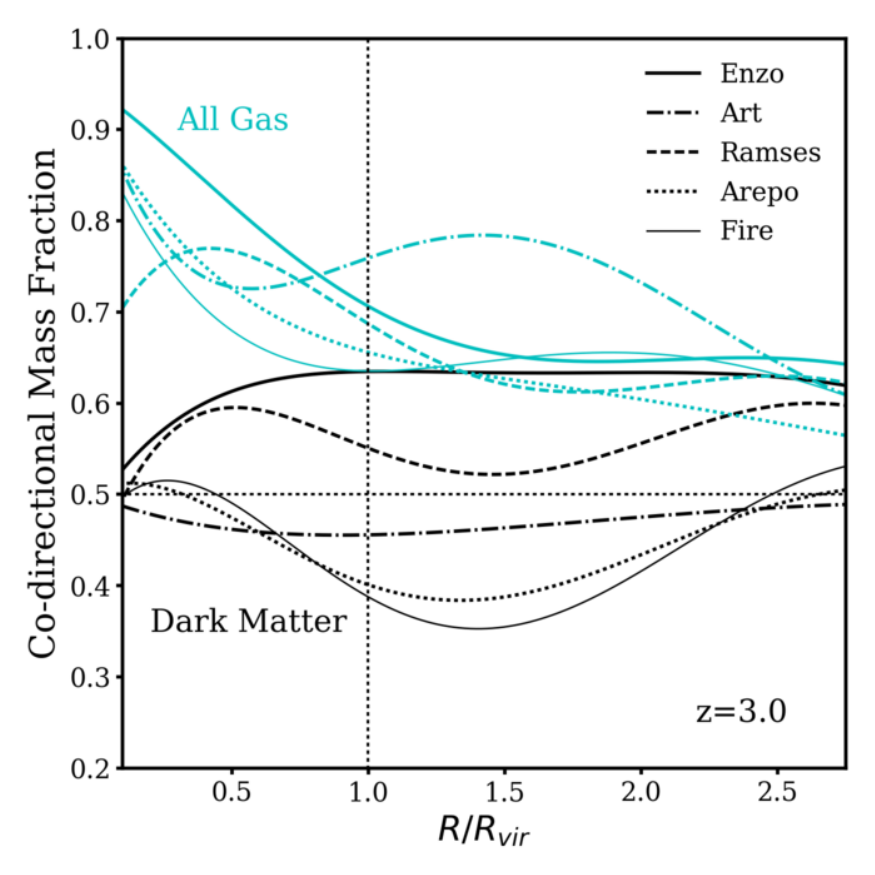}
 \includegraphics[width=0.34\textwidth]{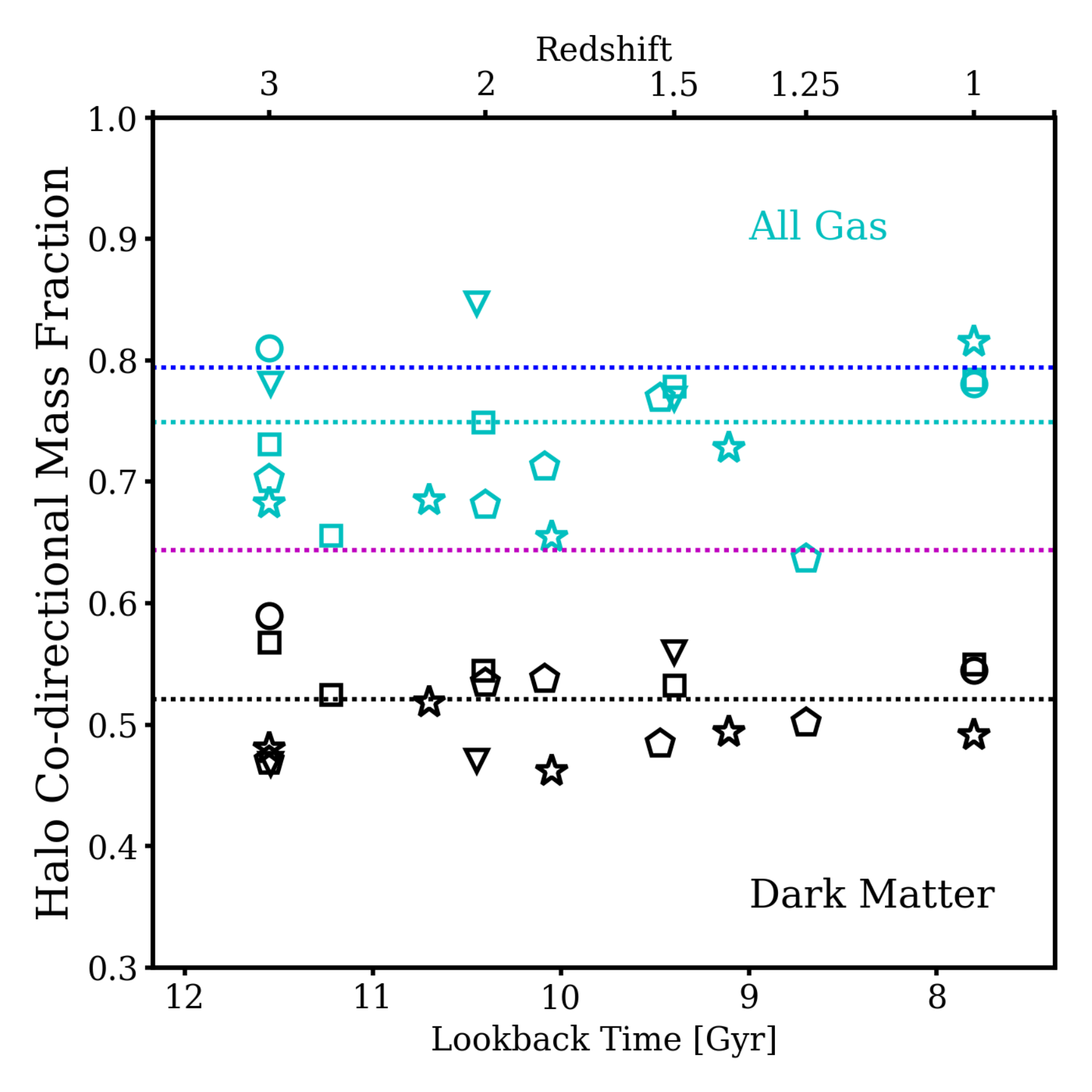}
 \includegraphics[width=0.34\textwidth]{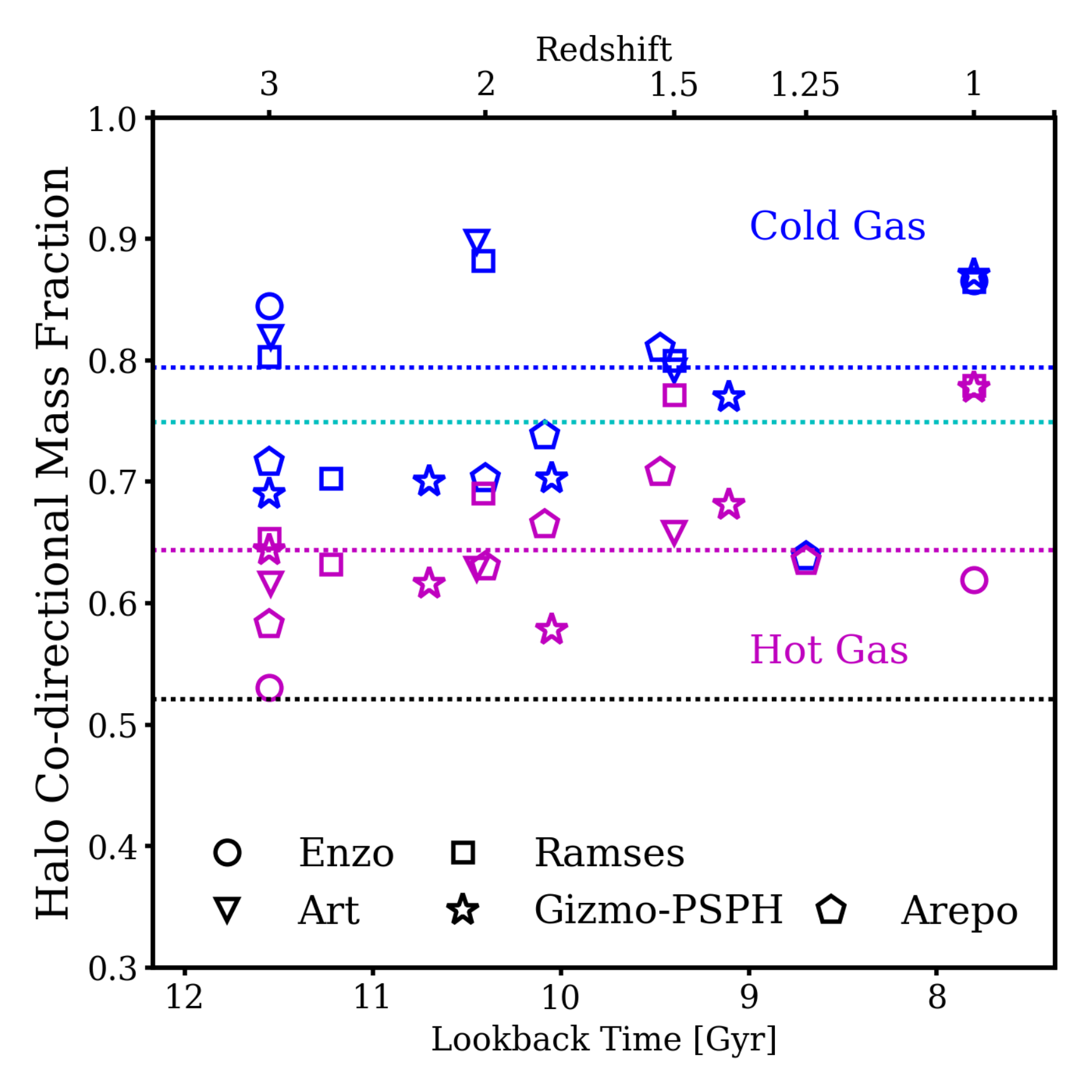}
 \caption{Left: co-directional mass fraction (see text for definition) of gas (cyan) versus dark matter (black) at $z=3$, as a function of radius.
                Middle/right: co-directional mass fraction in the halo ($0.1<R/R_{vir}<1.0$) for dark matter versus gas (middle panel) and for 
                cold versus hot gas (right panel).  The dotted horizontal lines in these panels represent averages over the entire redshift range and across all simulations for 
                all gas (cyan), cold gas (blue), hot gas (red), and dark matter (black).}
\label{fig_massfraction}
\end{figure*}
%>>>>>>>>>>>>>>>>>>>>>>>>>>>>>>>>>>>>>>>><<<<<<<<<<<<<<<<<<<<<<<<<<<<<<<<<

\subsection{Co-directional Halo Gas}
\label{corotation_massfrac}
As a means of quantifying this result, we define the ``co-directional mass fraction'' in the following way, at any given epoch.  For three arbitrary orthogonal projections,  
we define a cutting plane (passing through the center of the halo) that best divides the halo into positive versus negative line-of sight velocities for all gas within the virial radius.
Each particle (or cell, depending on code architecture) can then be defined as co-directional (along this projection) if its line-of-sight velocity was correctly categorized by this cutting 
plane.\footnote{If the halo gas were rotationally supported, we would say the gas is co-rotating, rather than co-directional; 
however we are hesitant to use this terminology, since the gas is actively spiraling inwards to the center of the halo, and co-rotation might be misinterpreted to imply angular momentum support.}
We then select the projection with the highest overall co-directional fraction for all gas in the halo 
(but not the galaxy: $0.1<R/R_{vir}<1.0$).
This selection typically corresponds to the projection in which the galaxy is seen closest to edge-on, though we note that this may not always be the case, if there 
is a significant misalignment between the angular momentum direction of the inspiraling gas and that of the galactic disk.  
The co-directional mass fraction of any given
component (dark matter, cold gas, hot gas, or all gas) is the mass fraction that has been categorized as co-directional along this preferred projection.

Using this definition, the left panel of Figure \ref{fig_massfraction} shows the co-directional mass fraction as a function of radius at $z=3$ (for all gas versus dark matter).
As might be expected, the average co-directional mass fraction for dark matter (taking into account all the simulations) is $\sim50\%$, though with considerable variation depending on 
the exact orientation of the co-directional cutting plane.
However, among all five simulations, the co-directional mass fraction for gas shows remarkably similar behavior with radius---declining smoothly from $\sim85\%$ 
at the galactic region ($R=0.1\Rvir$) to $\sim70\%$ at the virial radius.  Indeed, even extending to $\sim2.5\Rvir$, 
the co-directional mass fraction of gas in the cosmic web remains significantly higher than that of the dark matter, as expected if the line-of-sight velocity structure of 
the cold gas is kinematically linked to the filamentary gas flows beyond the virial radius of the halo.  

We explore the co-directional mass fraction over cosmic time in the middle and right panels of Figure \ref{fig_massfraction}, which use the same procedure outlined above to compute
a single value at each epoch for the total co-directional mass fraction in the halo (but not the galaxy: $0.1<R/R_{vir}<1.0$) for dark matter versus gas (middle panel), and further distinguishing 
between cold gas versus hot gas (right panel).  In both panels, the average for dark matter, all gas, cold gas, and hot gas among all simulations and over the entire redshift range 
from $z=3-1$ are given by the horizontal dotted lines.  
As shown in the figure, the dark matter co-directional mass fraction varies somewhat sporadically around $\sim50\%$, depending on the cutting plane orientation,
while the halo gas shows significantly higher co-directional mass fractions ($75\%$ for all halo gas).  
While there are significant variations in the detailed results among the simulations, with some codes showing stronger co-directional
kinematics than others, all simulations also demonstrate a higher tendency for cold gas to show this co-directional velocity structure in the halo over hot gas ($79\%$ versus $64\%$).

%>>>>>>>>>>>>>>>>>>>>>>>>>>>>>>>>> Gizmo merger<<<<<<<<<<<<<<<<<<<<<<<<<<<<<<<<<
\begin{figure*}[tbh!]
 \includegraphics[width=1.0\textwidth]{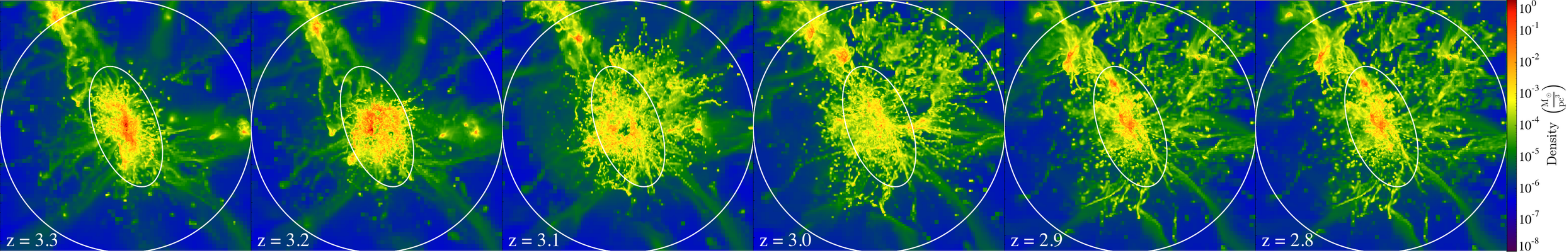}
 \includegraphics[width=1.0\textwidth]{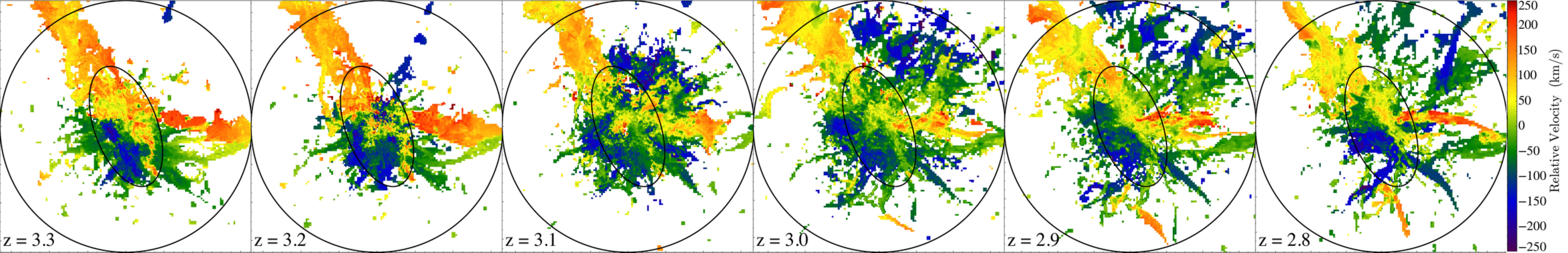}
 \caption{Time lapse of a post-merger violent outflow event in \gizmo$ $ at $z\sim3$, proceeding from left to right.  The overlaid circles denote the halo virial radius
 and the overlaid ellipse in each panel roughly corresponds to the region of coherent inspiraling gas at $z=3.3$, to aid the eye in comparison between images.
 Top: density map of the gas in the halo.  
 Bottom: line-of-sight velocity of cold dense gas in the halo (identical analysis to Figure \ref{fig_coldflowdisks}).  
  The coherent rotation in the bottom-left panel is effectively destroyed by the violent outflow from $z=3.2$---$3.0$, 
  but once the outflow event is over,
  fresh high angular momentum infall along the cosmic web begins to establish a new coherent inspiraling region by $z=2.8$, demonstrating the robustness of the inspiraling gas phenomenon.}
\label{fig_starburst}
\end{figure*}
%>>>>>>>>>>>>>>>>>>>>>>>>>>>>>>>>>>>>>>>><<<<<<<<<<<<<<<<<<<<<<<<<<<<<<<<<

Since the halo gas mass is dominated by its cold component (see Figure \ref{fig_mass_vstime}), 
it is worth exploring whether the above trend in co-directional 
mass fraction between the cold and hot components might be the result of an offset in angular momentum direction between the hot versus cold gaseous halos. 
After repeating the above analysis, but using the line-of-sight velocity of the \emph{hot} gas 
to define the co-directional cutting plane, we find only small variations in the above results.  
For example, the co-directional mass fraction in the halo at $z=3$ 
for [$\enzo, \art, \ramses, \arepo, \gizmo$] 
decreases slightly (if at all) for cold gas from [$84\%$, $82\%$, $81\%$, $72\%$, $69\%$] to [$79\%$, $81\%$, $81\%$, $72\%$, $69\%$], respectively, and 
increases slightly (if at all) for hot gas from  [$53\%$, $62\%$, $65\%$, $58\%$, $64\%$] to [$55\%$, $62\%$, $68\%$, $58\%$, $64\%$].  
Therefore, we conclude that the overall trends shown in Figure \ref{fig_massfraction}
are not highly sensitive to the way we define the co-directional cutting plane.

Taken together with our previous, more qualitative results, our findings suggest that 
across a broad range in hydrodynamic code types and subgrid physics models of galaxy formation,
the presence of co-directional inspiraling cold streams in galaxy halos at $z>1$
is a natural consequence of high angular momentum filamentary inflow along the cosmic web, and 
represents a robust prediction of cosmological gas accretion in LCDM.

\subsection{The Rapid Destruction and Re-formation of Coherent Inspiraling Gas at $z=3$}
\label{coldflow_regrow}

In the discussion of Figures \ref{fig_z3environment} and \ref{fig_coldflowdisks}, we noted that the velocity structure of the galaxy in the \gizmo$ $ simulation
is not nearly as clean and orderly as the other simulations at $z=3$, and therefore does not seem to host the same 
clear co-directional velocity structure.
While we thought it important to show all galaxies at precisely the same epoch,
we note that in \gizmo, the galaxy happens to be in the midst of a post-merger starburst, accompanied by a violent outflow event at this epoch, due to this code's strong 
feedback physics. Halo-halo mergers, of course, tend to occur at broadly similar times in all codes, but differences in galaxy masses and halo baryonic mass distributions mean that the galaxy-galaxy 
mergers can and do occur at significantly different times, and with different mass ratios and corresponding consequences for star formation, at the 
halo center \citep[see, e.g.][]{Stewart09b,Hopkins10dec}.
The obvious clumpiness of the outflows may owe, at least partially, to well-known numerical difficulties capturing fluid-mixing instabilities in SPH, 
even in the improved P-SPH implementation; this is supported by early results from the FIRE-2 simulations, which use a different, mesh-free Godunov-type finite volume method to solve the 
hydrodynamics (P. Hopkins, private communication).

In Figure \ref{fig_starburst}, we show the structure of this galaxy immediately before and after this violent merger event.  The time sequence begins in the 
left panel at $z=3.3$, where inflowing cold gas demonstrates 
the same line-of-sight velocity structure as in Figure \ref{fig_coldflowdisks},
including the presence of co-directional inspiraling cold streams, which have initially taken the apparent  form of an extended disk-like structure 
reminiscent of the ``cold flow disks'' reported previously in cosmological simulations \citep[e.g.][]{Stewart11b}.
To aid the eye in comparing the images, some of 
which are quite chaotic during the outflow, an identical ellipse has been overlaid on each image roughly corresponding to this coherent inspiraling gas region.
At this epoch, the recent influx of 
of fresh gas onto the central regions of the galaxy results in a spike in star formation, and consequently a 
violent spherical outflow event from $z=3.2$---$3.0$ that effectively destroys the ISM of the galaxy (leaving a deficit of gas in the center of the galaxy, as seen at $z=3.1$) 
and disrupting the inflowing filamentary gas in the CGM of the halo.  
However, the filamentary gas continues to flow into the halo, and this inflow continues to contribute substantial angular momentum.  As a result, 
a new co-directional velocity structure becomes apparent
almost immediately after the outflow event has subsided, with the co-directional mass fraction of cold dense gas in the halo (defined as described in \S\ref{corotation_massfrac})  changing rapidly 
from $81\%$ at $z=3.3$ (before the outflow)
to $\sim60\%$ during the outflow, back up to $76\%$ at $z=2.8$ (post-merger).
By the rightmost panels, the co-directional inspiraling cold streams have once again formed a roughly disk-like structure, 
along a very similar orientation to the original inspiraling gas structure.
We argue that the bursty nature \citep{Muratov15} of the subgrid physics as implemented in \gizmo$ $ coupled with this demonstration of the near-immediate regrowth of 
the coherent inspiraling gas structure 
after a massive outflow event only reinforces the robust nature of 
inspiraling cold streams in the halos of massive galaxies in LCDM.

\section{Discussion}
\label{discussion}
Past studies of galaxy formation simulations have reported the existence of co-rotating structures of cool gas in the outskirts of galaxy halos.  
Our results indicate that inspiraling halo gas of this kind is robust to different feedback models and  hydrodynamic solvers.   
One implication of this result is that extended, high-angular momentum cold stream configurations offer a testable observational 
signature of the LCDM galaxy formation paradigm.
For example, the fact that a large fraction of the halo gas have velocities that are co-directional means that observations of the gas (e.g. from quasar absorption systems) 
will show a blueshifted and redshifted side that is usually interpreted as rotation.  We emphasize that in this case most of the gas is far from being angular momentum 
supported and that ``inspiraling'' halo gas is a more accurate description of its coherent motion than ``rotating''.

Encouragingly, there is a growing body of observational evidence that seems to indicate that co-directional halo gas is indeed seen around real galaxies. 
For example, kinematic studies of some Ly$\alpha$ nebulae suggest rotational velocities and inflow rates consistent with those expected for these inspiraling streams
\citep{Martin14,Prescott15}.  Similarly, 
absorption line studies are beginning to emphasize the bimodal distribution of absorption detections, where detections along the galaxy's 
minor axis tend to show absorption properties consistent with outflowing gas, while detections roughly along the galaxy's major axis 
demonstrate properties (such as co-rotational inflow) that are consistent with inspiraling cold streams 
\citep{Kacprzak10,Kacprzak12a,Kacprzak12b,Bouche12,Bouche13,Crighton13,Nielsen15,Bouche16,Ho17}

Perhaps the most direct confirmation of the existence of inspiraling halo gas comes from 
\cite{Martin15}, who performed a spectroscopic analysis on the cosmic filament (illuminated by two nearby QSOs) first detected by \cite{Cantalupo14} at $z\sim2$.
They found that a substantial fraction of the illuminated region was in fact a huge co-rotating gaseous structure. 
The extremely extended gaseous ``disk'' (extending to $\sim\Rvir/2$, corresponding to a width of $125$ physical kpc) showed smooth rotation kinematics,
with one side of the disk kinematically linked to the inflow velocity of the nearby cosmic filament.  
This very closely resembles 
what we have presented here for co-directional inspiraling cold streams, though we note that the particular system observed by \cite{Martin15} was 
estimated to be a much more massive halo than what we have simulated here ($\Mvir\sim10^{13}\Msun$) and it therefore reported a correspondingly more 
massive and extended protogalactic disk than found in our simulations, as might be expected for a larger, more massive halo.  
A similar cold flow protodisk, again fed by a cosmic filament that was first detected in Ly$\alpha$ emission, was also 
reported in \cite{Martin16}, suggesting that inspiraling disk-like structures may be common phenomena for massive galaxies at high redshift.

While not seen in our particular simulations, we also speculate that 
polar ring galaxies---which have previously been suggested as evidence of cold flow gas accretion 
onto galaxies \citep{Maccio06,Brook08,Spavone10}---may be a result of a similar phenomenon.
Such galaxies could reasonably occur when strong central torques (e.g., from a major galaxy merger) result in a
near perpendicular misalignment between the angular momentum 
of the central galaxy and that of the inflowing cold mode gas.  

We note that the inspiraling cold streams in our simulations are significantly more massive and extended (relative to the halo virial radius) at high redshift,
when cosmic filaments are more narrowly defined and contain higher density gas flows.  However, Figure \ref{fig_spin_histogram} and 
previous work \citep[e.g.,][]{Stewart13} both demonstrate that accreting cold gas continues to have high angular momentum, even at later times where the 
rotational signature of a continuous gaseous structure may be less clear.  
Additionally, Figure \ref{fig_massfraction} demonstrates that the co-directional mass fraction of cold gas in the halo stayed consistently high over cosmic time
(at least until $z=1$).
We speculate that it may be possible that this high angular momentum accretion helps to explain observations of 
extended XUV disks \cite[e.g.][]{Thilker05,Thilker07,Lemonias11,Holwerda12}, local extended \HI$ $ disks 
\citep[e.g.][]{GarciaRuiz02,Oosterloo07,Walter08,ChristleinZaritsky08,Sancisi08,Wang13,Huang14,Courtois15},
and co-rotating cold halo gas around local Milky Way analogs \citep[e.g.][]{Diamond-Stanic15}.  

Indeed, these growing observations of high angular momentum material in the outskirts of galaxy halos
would be quite difficult to explain if one were to assume the canonical picture of galaxy formation whereby baryons in galaxy halos 
share the same distribution of angular momentum as the dark matter.  In contrast, the cold flow paradigm naturally predicts that halo gas (and particularly the cold halo gas) 
preferentially constitutes recent gas accretion from the cosmic web, with $\sim3$---$5$ times the angular momentum of the dark matter, naturally explaining
the kinds of high angular momentum phenomena being observed.  
We caution, however, that we have not focused on the gaseous
halos in our simulations at $z<1$ here, and leave a more detailed comparison between simulations and low-z observations as a topic of further study.

\section{Conclusion}
\label{conclusion}
We have simulated the evolution of a Milky Way-sized galaxy from identical
cosmological initial conditions with a variety of simulation codes: \enzo, \art, 
\ramses, \arepo, and \gizmo.  Each code has used subgrid physics models drawn from scientific literature common to 
each simulation type, and we have compared the simulations in an attempt to draw robust conclusions about galaxy formation in 
LCDM (focusing on $z>1$) that are not sensitive to uncertain aspects of galaxy formation simulations.  To ensure uniform analysis 
among the various code types, we have used the hydrodynamic analysis software \yt, which enables the same analysis routine to be performed 
on each code.

While we found many aspects of the simulated galaxies that did vary substantially among the simulations (e.g. morphology, stellar mass, hot gas halo temperature, and mass, to 
be discussed further in future work), we found the following qualitative features common to all simulations, regardless of which subgrid physics model or hydrodynamic code was used:

\begin{enumerate}

   \item Gas in the galaxy halo has substantially higher specific angular momentum than the dark matter in the halo, with mean values of 
     $j_{\rm cold}\simeq 4 j_{\rm DM}$ and $j_{\rm hot}\simeq 2 j_{\rm DM}$ (though with considerable scatter), leading to a typical cold halo gas spin parameter of $\lambda_{\rm cold}\simeq 0.12$.

  \item  The large-scale filamentary structure is qualitatively similar in all simulations (with minor variations, for example regarding lower mass streams of secondary importance 
    to the galaxy's growth).
    The three-dimensional geometry of these filaments, which are all flowing toward the central galaxy (the highest overdensity in its environment) 
   results in a strong line-of-sight velocity structure.  Filaments flowing onto the galaxy 
   from opposite directions (along an arbitrary line-of-sight) tend to show alternating blueshifted and redshifted velocities relative to the galaxy as they flow toward the galaxy center.

  \item  As the filamentary gas accretion enters the virial radius, the large-scale velocity 
    structure of the accreting filaments inevitably  results in inspiraling cold streams in the halo of the galaxy that carry significant angular momentum 
    as they spiral in from the virial radius to the galactic region.
     For the Milky Way-size halo simulated here, the maximum line-of-sight velocity expected for these inspiraling streams is 
     $\sim250$ km/s (corresponding to roughly $1.5$ times the virial velocity of the halo).
     As a result, the vast majority ($\sim80\%$) of cold halo gas follows a clear co-directional velocity structure 
    (with a single cutting plane dividing positive versus negative line-of-sight velocities) as the cold streams spiral toward the center of the halo.  
    In contrast, the co-directional 
    mass fractions in the halo are considerably lower for the hot gas ($\sim65\%$) or the dark matter ($\sim50\%$). 
    
    \item
    Inspiraling cold streams occasionally  take the previously reported morphology of cold flow disks: high angular momentum cold gas that is transitioning 
    from the cosmic web, though the halo as a roughly disk-like structure (except that the gas is not angular momentum supported) and will eventually accrete onto the galactic disk.    
    These coherent inspiraling structures represent continuous and dynamic flows from the cosmic web;
    even after a violent outflow event disrupts the CGM in one of the simulations, the newly inflowing gas 
    rapidly re-forms a similar inspiraling structure 
    along a the same orientation soon after the outflow has subsided.
    
\end{enumerate}

In this work, we have limited our analysis to the growth of a single Milky Way-size halo at $z>1$ using a variety of different hydrodynamic codes and feedback physics implementations. 
It is therefore difficult to draw general conclusions about galaxy formation from the simulation of a single halo; however, a number of theoretical works have previously established the 
high angular momentum nature of filamentary gas accretion, using various hydrodynamic codes, larger cosmological volumes, and/or analysis of multiple zoom-in simulations.  
For example, \cite{Pichon11} analyzed $\sim15,000$ halos at $z>1.5$ from a (lower resolution) cosmological-scale simulation using the \ramses$ $ code; 
in a companion work, \cite{Kimm11} also included $\sim900$ intermediate resolution halos and two high-resolution zoom-in simulations to $z=0$ using \ramses; 
\cite{Stewart11b,Stewart13} analyzed $4$ zoom-in simulations to $z=0$ using the SPH code \gasoline; 
and \cite{Danovich15} analyzed $29$ zoom-in simulations at $z>1.5$ using the \art$ $ code.  
The results presented here demonstrate that the high angular momentum nature of cold gas accretion in LCDM is not likely to change (in the qualitative sense)
among a broad range of different physics implementations and hydrodynamic codes, suggesting that 
the presence of inspiraling cold streams in galaxy-size halos 
appears to be a robust expectation of LCDM.

However, we note that there are considerable variations in the
quantitative nature (morphology, rotational velocity, size, temperature, density, etc.) of the inspiraling cold streams in each of the simulations performed in this work.
The detailed properties and prevalence of these inspiraling streams 
are yet to be fully understood, and cannot be determined from the single high resolution simulation presented here.  
The co-directional velocity structure noted here is also likely to depend on the geometry and kinematics of 
the cosmic web in the galaxy's environment, so we speculate that there are likely to be significant environment effects, even at fixed halo mass.  
For example, Milky Way-sized galaxies near the outskirts of galaxy clusters would not be expected to
dominate the gravitational potential of the cosmic web in their large-scale environment, 
so we may not expect to find the same clear co-directional velocity signature for filamentary inflow (i.e., Figure \ref{fig_z3environment}) for such systems either.

We also take special note that in several of the simulation codes used here (and in previous works), these inspiraling streams result in the formation of 
transient inspiraling disk-like structures qualitatively similar to the ``cold flow disks'' of, e.g., \cite{Stewart13}.
The exploration of the prevalence of these inspiraling gaseous structures in simulations, for different environments and halo masses, 
would be a useful topic of further study, especially in light of 
recent observations of large co-rotating gaseous structures at $z\sim2$ \citep{Martin15, Martin16} that are strikingly similar to the qualitative results presented here.

\acknowledgements
Computations described in this work were performed using the publicly available \enzo$ $ code (http://enzo-project.org), and the publicly available \yt$ $ toolkit (http://yt-project.org/), 
both of which are the products of collaborative efforts of many independent scientists from numerous institutions around the world.  
Their commitment to open science has helped make this work possible.
KRS would especially like to thank Matt Turk and Nathan Goldbaum for their support with \yt.  
KRS also thanks Volker Springel, Lars Hernquist, and Paul Torrey for running the \arepo$ $ simulation discussed in this work, for allowing us to include its results as part of this comparison project, and for providing useful discussions and comments.

KRS was supported by HST-GO-14268.026-A.  JSB was supported by HST AR-12836. 
CAFG was supported by NSF grants AST-1412836 and AST-1517491, by NASA grant NNX15AB22G, and by STScI grants HST-AR-14293.001-A and HST-GO-14268.022-A.
JD acknowledges support from the Spin(e) grant ANR-13-BS05-0005 of the French Agence Nationale de la Recherche (http://cosmicorigin.org).
Support for PFH was provided by an Alfred P. Sloan Research Fellowship, NASA ATP Grant NNX14AH35G, and NSF Collaborative Research Grant \#1411920 and CAREER grant \#1455342.
%art
The \art$ $ run was performed
at the National Energy Research Scientific Computing Center (NERSC) at  
Lawrence Berkeley National Laboratory.
DC acknowledges support from the European Research Council under the European Community’s Seventh 
Framework Programme (FP7/2007-2013) via the ERC Advanced Grant ‘STARLIGHT: Formation of the First Stars’ (project number 339177).
%ramses
The \ramses$ $ simulation was performed on the DiRAC Facility, jointly funded by BIS and STFC.
JD’s research  is partly funded  by Adrian Beecroft and the Oxford Martin School.
%gsph
%hpc
The \gizmo$ $ run was carried out on the CUNY HPC which is supported, in part, under National Science
Foundation Grants CNS-0958379, CNS-0855217, ACI-1126113, and the City University of New York High Performance Computing Center at the College of Staten Island.
%ctp
Storage and analysis of the simulations was performed on the CTP cluster, which is supported by
GRTI grant CT04AGR15001 and the Physics Division of the U.S. Army Research Office grant $\#$64775-PH-REP.
\bibliography{scylla1}

\end{document}